\documentclass[showpacs,preprint,prd,12pt]{revtex4-1}

\usepackage{graphicx} % This is already loaded by the atlasnote class
\usepackage{amsmath}
\usepackage{subfigure}
\usepackage{epstopdf}
\usepackage{multirow}
\usepackage{xspace}
\usepackage{rotating}
\usepackage{longtable}
\usepackage{multirow}
\usepackage[breaklinks=true]{hyperref}
\hypersetup{
  colorlinks=true,
  linkcolor=blue,
  citecolor=blue,
  urlcolor=blue
}

\graphicspath{{./fig/}}
\usepackage{array,tabularx,epsfig,mathrsfs,graphicx,rotating}
\usepackage{ifthen}
\usepackage{amsfonts}
\newcommand{\beq}{\begin{equation}}
\newcommand{\eeq}{\end{equation}}

\chardef\til=126

\newcommand{\fbb}{\mathrm{fb^{-1}}}
\newcommand{\abb}{\mathrm{ab^{-1}}}
\newcommand{\mjj}{M_{jj}}
\newcommand{\pythia}{{\sc Pythia8}\xspace}

\begin{document}

\preprint{ANL-HEP-139751}
\date{\today}

%%%%%%%%%%%%%%%%%%%%%%%%%%%%%%%%%%%%%%%%%%%%%%%%%%%%%%%%%%%%%%%
\title{
Precision searches in dijets  at the HL-LHC and HE-LHC
}
%%%%%%%%%%%%%%%%%%%%%%%%%%%%%%%%%%%%%%%%%%%%%%%%%%%%%%%%%%%%%%%

\author{S.~V.~Chekanov, J.~T.~Childers, J.~Proudfoot, R.~Wang}
\affiliation{
HEP Division, Argonne National Laboratory,
9700 S.~Cass Avenue, Argonne, IL 60439, USA 
}% 

\author{D.~Frizzell}
\thanks{Also affiliated with the HEP Division, Argonne National Laboratory,
9700 S.~Cass Avenue, Argonne, IL 60439, USA
}
\affiliation{
Homer L. Dodge, Department of Physics and Astronomy, University of Oklahoma, Norman, OK, USA
}%

%\hfill\draft{Version 1.6}

\begin{abstract}
This paper explores the physics reach of 
the High-Luminosity Large Hadron Collider (HL-LHC) for searches 
of new particles decaying to two jets. We discuss inclusive searches in dijets and $b$-jets, as well
as searches in semi-inclusive events by requiring an additional lepton
that increases sensitivity to different aspects of the underlying processes. 
We discuss the expected exclusion limits for generic models predicting 
new massive particles that result in resonant structures in the dijet mass. 
Prospects of  the Higher-Energy LHC (HE-LHC) collider are also discussed.
The study is based on  the  Pythia8 Monte Carlo generator  using
representative event statistics for the HL-LHC and HE-LHC running conditions.
The event samples were created using  
supercomputers at NERSC.
\end{abstract}


\maketitle

%%%%%%%%%%%%%%%%%%%%%%%%%%%%%%%%%%%%
%            Content               % 
%%%%%%%%%%%%%%%%%%%%%%%%%%%%%%%%%%%%

%\input{sect_intro.tex}
%%%%%%%%%%%%%%%%%%%%%%%%%%%%%%%%%%%%%%%%%%%%%%%%%%%%%%%%%%%%%%%%%%
\section{Introduction}
%%%%%%%%%%%%%%%%%%%%%%%%%%%%%%%%%%%%%%%%%%%%%%%%%%%%%%%%%%%%%%%%%%
\label{sec:intro}

Heavy particles decaying 
to two jets are a generic consequence of many Beyond-the-Standard Model (BSM) theories.
Recently, searches in dijet mass distributions at the LHC have been performed by 
both ATLAS and CMS collaborations \cite{Aad:2015xis,2016229, Aaboud:2017yvp, Khachatryan:2010jd, Chatrchyan:2013qha, Sirunyan:2016iap}  
using run I and run II LHC data.
An extension of such studies  at the High-Luminosity LHC (HL-LHC) and
the Higher-Energy LHC (HE-LHC) colliders will be an important  goal of 
the energy frontier. 

In the past, searches in dijets  at the LHC were  mainly focused on the 
high-mass tail of dijet distributions, rather than
on the bulk of data (below 1~TeV).  
This is related to the fact that the collection of inclusive dijet events with dijet masses of the order of 
hundreds of GeV is reduced (or ``pre-scaled'') at the trigger level in order to cope with a large rate of multijet QCD events. 
Focusing on the tail of dijet-mass distributions ($\mjj$),
rather than on the data in the region close to the electroweak (EWK) scale, $<1$~TeV, 
limits the potential of searches in inclusive dijet masses below 1~TeV. 

Requiring a lepton, photon and or other identified objects  with relatively low transverse momentum ($p_T<0.1$~TeV), 
allows searches for new particles through associated production in dijet masses using the main fraction of collected data.
This can lead to  detailed
studies of dijets at relatively low masses ($0.1<\mjj<1$~TeV),
while reducing  contributions from QCD multijet events,
which represent the main background for inclusive dijet mass searches.
The multijet  QCD background can further be reduced for dijets
where one or two jets are tagged as $b-$jet. 
High-precision searches focusing on the
medium range of invariant masses in semi-inclusive final states was 
discussed in \cite{stalk} in the context of
a broad class of Hidden Valleys models \cite{Strassler:2006im}, 
in which new particles can be as light as the Standard Model (SM) particles, i.e. with masses below 1~TeV.  

The studies of dijets involve many technical challenges for future experiments. 
The HL-LHC will deliver about $3~\abb$ of integrated luminosity, more than a factor 10 of the data 
that will be collected by the end of the LHC project.
This amount of data  opens a new chapter in BSM searches focusing on extraction of 
features in $\mjj$ distributions
where the relative statistical uncertainty (i.e. statistical uncertainty 
expressed as a fraction of data point value)  can be as small as $10^{-4}$ -- $10^{-3}$ 
near $\mjj=0.1$~TeV as shown in this paper. 
As a first step to such studies, we will use Monte Carlo (MC) event generation
with a representative event statistics 
to explore sensitivity to new states decaying to two jets, without 
simulations of detector effects and pile-up contributions.
%We hope that features in the form of relatively narrow peaks from heavy states 
%with $\Gamma/M<0.2$ still can be detectable. Therefore,  an exploration of the question of 
%realistic event statistics and its impact on such high-precision searches in  $M_{jj}$ distributions
%can be worth pursuing goal. As a first step to such studies, we will use Monte Carlo event generation without 
%simulation of detector effects and pile-up contributions.

The goal of this paper is to understand the physics potential 
of the proposed HL-LHC and HE-LHC experiments with respect to searches for new states
decaying to dijets for inclusive and semi-inclusive event selections, as well as  
to explore different methods designed for gaining more sensitivity to new physics.
In the searched mass region from $\sim 0.1$ to 10 (20)~TeV at the HL-LHC (HE-LHC),
event rates fall by more than 14 orders of magnitude.
Modeling mis-identification rates of leptons and $b-$jets in this large range of invariant masses
is challenging since it requires certain experiment-motivated  assumptions
that can numerically be implemented in MC simulations,
as well as an analysis of large event samples from MC generators that include parton showers and hadronization.
We will discuss this topic using realistic MC event samples, creation of which
has become possible with the use of high-performance computing.
In addition, we will calculate the exclusion limits  
for BSM models predicting heavy particles decaying to two jets at  
the HL-LHC and HE-LHC. The limits will be calculated using MC simulations of  
inclusive dijets, $b$-jets and dijets associated with a lepton.

%\input{sect_mc.tex}
%%%%%%%%%%%%%%%%%%%%%%%%%%%%%%%%%%%%%%%%%%%%%%%%%%%%%%%%%%%%%%%%%%
\section{Monte Carlo event simulations}
\label{sect_mc}
%%%%%%%%%%%%%%%%%%%%%%%%%%%%%%%%%%%%%%%%%%%%%%%%%%%%%%%%%%%%%%%%%%

The analysis presented in this paper was performed using the \pythia \cite{Sjostrand:2006za, *Sjostrand:2007gs}
MC generator with the default parameter settings and the ATLAS A14 tune \cite{ATL-PHYS-PUB-2014-021} for minimum-bias events.
The center-of-mass collision energy of $pp$ collisions 
was set to 14~TeV and 27~TeV for the HL-LHC and HE-LHC respectively.
The  NNPDF 2.3 LO \cite{Ball:2012cx, *Ball:2014uwa} parton density function, 
interfaced with \pythia via the LHAPDF library \cite{Buckley:2014ana},  was used.
A minimum value of transverse momentum  for the matrix elements for  $2\rightarrow 2$ processes
was 40~GeV. The simulations  were created for three categories of SM  
processes implemented in leading-order (LO) matrix elements,  with the parton shower (PS) followed 
by hadronization:

\begin{itemize}

\item
Light-flavor QCD dijets. 
This category of events includes  ten $2\rightarrow 2$ quark and gluon processes, including $b$-quark pair production,
but excluding $t\bar{t}$ production from hard interactions, which is considered as separate below.  We apply a phase-space re-weighting to
increase the statistics in the tail of the $M_{jj}$ distribution as discussed in \cite{Sjostrand:2006za, *Sjostrand:2007gs}.

\item
Vector and scalar boson production that includes  the $W$  $Z$, $H^0$-boson processes available in \pythia. 
This category of events has  23 processes at the $2\rightarrow 1$  and $2\rightarrow 2$ level.
Due to the presence of the $2\rightarrow 1$ processes, no phase space re-weighting was used.

\item
$t\bar{t}$  and single top quark production which includes six $2\rightarrow 2$ processes. No phase-space re-weighting was applied.

\end{itemize}

Currently, we do not use simulations at next-to-leading-order (NLO) accuracy, or 
at tree-level LO matrix elements included in 
{\sc Alpgen}~\cite{Mangano:2002ea} or {\sc Blackhat}~\cite{Bern:2012my},  which
typically lead to larger cross sections. 
Even with the use of supercomputers, the large data samples required for the 
statistical precision of this study preclude using these more computationally 
intensive programs.

The simulation tools, small portions of event samples and \pythia settings  used in this study are available
from the HepSim public event repository \cite{Chekanov:2014fga}.
The studies presented in this paper, however,
require significant statistics, thus keeping events on a disk is impractical.
A faster and less storage-demanding solution based on generating  parton-level events 
does not provide the required information,   
since the studies presented in this  paper are based on distributions sensitive
to mis-identification of light jets as leptons or $b$-jets. An analysis   
of large event samples from the complete simulation of the parton shower followed by  hadronization  is essential. 
A simple scaling of low-statistics distributions 
to a luminosity of the order of several
$\abb$ leads to significant fluctuations in $\mjj$ bins, even after  the  phase space re-weighting used in this paper. 

In view of the above difficulties, 
we chose to perform the generation and analysis in series on a supercomputer.
To achieve this, a Docker/Singularity container image of
the HepSim software was created and deployed on 
the Cori supercomputer (Phase 2, Intel Knights Landing cores) of  
the  National Energy Research Scientific Computing Center (NERSC).  
The software image includes the \pythia MC generator and the complete HepSim software stack for jet reconstruction 
and the final analysis. The maximum value of the \pythia seed ($9\cdot 10^{8}$) is not very high 
for massively parallel jobs, therefore,  
special care was taken to avoid creation of duplicate events.  The calculations  
took about 10 million core-hours over a ten day period. 
The analysis used  
about 100 billion MC events for the three process categories discussed above at
the centre-of-mass energies of 14~TeV and 27~TeV.

The simulated number of background events for $\mjj<1$~TeV is several orders of magnitude smaller than
what is needed for the analysis of $\mjj$ distributions from the HL-LHC.  
For example, the cross section for 
multijet events with the 40~GeV cut on the LO matrix elements for  $2\rightarrow 2$ processes
in $pp$ collisions at 14~TeV  is $7.3\cdot 10^{10}$~fb. 
The problem of low statistics is 
partially  mitigated  using the adopted phase-space re-weighting technique discussed above.
As for any counting experiment the statistical uncertainty for each $\mjj$ bin was defined
as the square root of the bin height.

For the  simulation of the signal events, we use  a model with 
an extra gauge boson, $Z'$, that arises in many extensions \cite{Langacker:2008yv} 
of the electroweak symmetry of the Standard Model. 
The signal events were generated using the \pythia generator with the default settings, 
ignoring interference with SM processes, with 
a width of about $15\%$ of the $Z'$ mass.
This width is usually considered as the maximum width of a resonance in ATLAS searches \cite{Aad:2015xis,Aaboud:2017yvp}. 
We consider two cases for $Z'$ decays, one in which $Z'$ decays to light-flavor quarks, and one 
with only  $b$-quark decays.

% referee's comment
As discussed in the introduction, pile-up events were not included in the simulation. Mixing billions
of generated events with 140-200
low-$p_T$ ``minimum bias'' events is beyond the technical capability of
the computational resources used for this work.
We should note, however, that pile-up events are not expected to change
our conclusions  related  to  high-$p_T$ physics above the TeV scale.
In addition, the LHC experiments have developed
successful techniques to mitigate pile-up effects. Such techniques should be considered
in conjunction with detector-level objects (such as calorimeter clusters,  or tracks
associated with the primary interaction vertex), all of which are beyond the scope of this paper.

%%%%%%%%%%%%%%%%%%%%%%%%%%%%%%%%%%%%%%%%%%%%%%%%%%%%%%%%%%%%%%%%%%
\section{Event reconstruction}
\label{sec:reco}
%%%%%%%%%%%%%%%%%%%%%%%%%%%%%%%%%%%%%%%%%%%%%%%%%%%%%%%%%%%%%%%%%%

Hadronic jets were reconstructed from stable particles, which are
defined as having a lifetime more than $3\cdot 10^{-10}$ seconds. 
Neutrinos were excluded from consideration.
The jets were
reconstructed using the anti-$k_T$ algorithm \cite{Cacciari:2008gp} as implemented in the {\sc FastJet} package~\cite{Cacciari:2011ma}.
The jet algorithm used a distance parameter of $R=0.4$.
The minimum transverse momenta of jets was $40$~GeV, and the pseudorapidity of jets was $|\eta|<2.4$.

The minimum transverse  momentum of the leptons used in this analysis was set to 60~GeV.
To reduce the mis-identification rates, the leptons are required to be isolated.
A cone of the size $0.2$ in the azimuthal angle and pseudo-rapidity is 
defined around the true direction of the lepton. Then, all energies of particles inside this cone are summed. 
A lepton is considered to be isolated of it carries more than $90\%$ of the cone energy.  

As full simulation is not within the scope of this study, 
we estimate the rate of misidentification of muons (the muon fake rate) as a fraction, 
0.1\%, of the jet rate similar to the ATLAS study \cite{Aad:2009wy}. 
This is implemented by assigning the probability of $10^{-3}$ 
for a jet to be identified as a muon using a random number generator. 
We do not use electrons
since their fake rate is a factor of ten larger than for muons.

Dijet invariant masses, $\mjj$,  were reconstructed by combining the two leading jets
having the highest $p_T(jet)$. The minimum value of $\mjj$ was chosen to be 125~GeV, which is large enough to avoid biases arising from the 
jet selection and  contributions from the $W/Z$ decays. 
At the same time, this value well represents the bulk of the anticipated HL-LHC data 
where the $\mjj$ distribution smoothly decreases with increase of $\mjj$. This feature is  
important for our discussion in Sect.~\ref{sect_signal}.
%Note that, below $\mjj=125$~GeV, the distribution of $\mjj$ starts to decrease when using events
%triggered by realistic LHC triggers based on   
%leptons (dijets associated with a lepton will be considered in Sect.~\ref{sect_dijets_mu}).

A reproduction of the experimental mis-tag rate of $b$-jets 
is difficult for the generator-level MC studies.
Generally, the mis-tag rate depends on many factors, including a dependence on $p_T(jet)$ \cite{Aad:2015ydr}.
For the studies involving $b$-jets, we assume a constant 10\% mis-tag rate. 
This value is sufficiently realistic \cite{Aad:2015ydr} 
for large $p_T(jet)$ considered in this paper and, at the same time,
can easily be reproduced or modified in future studies.  
This  mis-tag rate was implemented by using the probability of $0.1$ for a light jet to be identified as a $b$-jet. 
The $b$-jets are additionally selected by requiring: (1) the $\Delta R$ between the $b$-quark and jet to be less than 0.4;
(2) the $b$-quark $p_{T}$ is at least 50\% of the jet $p_{T}$.

This paper, being based on the generator-level Monte Carlo samples, 
does not include simulations of detector efficiencies for reconstruction of jets, leptons and $b-$tagging.
The inclusion of efficiencies may change the exclusion limits shown in this paper, but they are unlikely to  change
comparisons between the HL-LHC and HE-LHC scenarios.

%%%%%%%%%%%%%%%%%%%%%%%%%%%%%%%%%%%%%%%%%%%%%%%%%%%%%%%%%%%%%%%%%%
\section{Dijets in inclusive events}
\label{sect_dijets}
%%%%%%%%%%%%%%%%%%%%%%%%%%%%%%%%%%%%%%%%%%%%%%%%%%%%%%%%%%%%%%%%%%

%\subsection{Expectations for the LHC and HL-LHC}

\begin{figure}[h]   
\begin{center}
   \subfigure[ $100~\fbb$] {
   \includegraphics[width=0.45\textwidth]{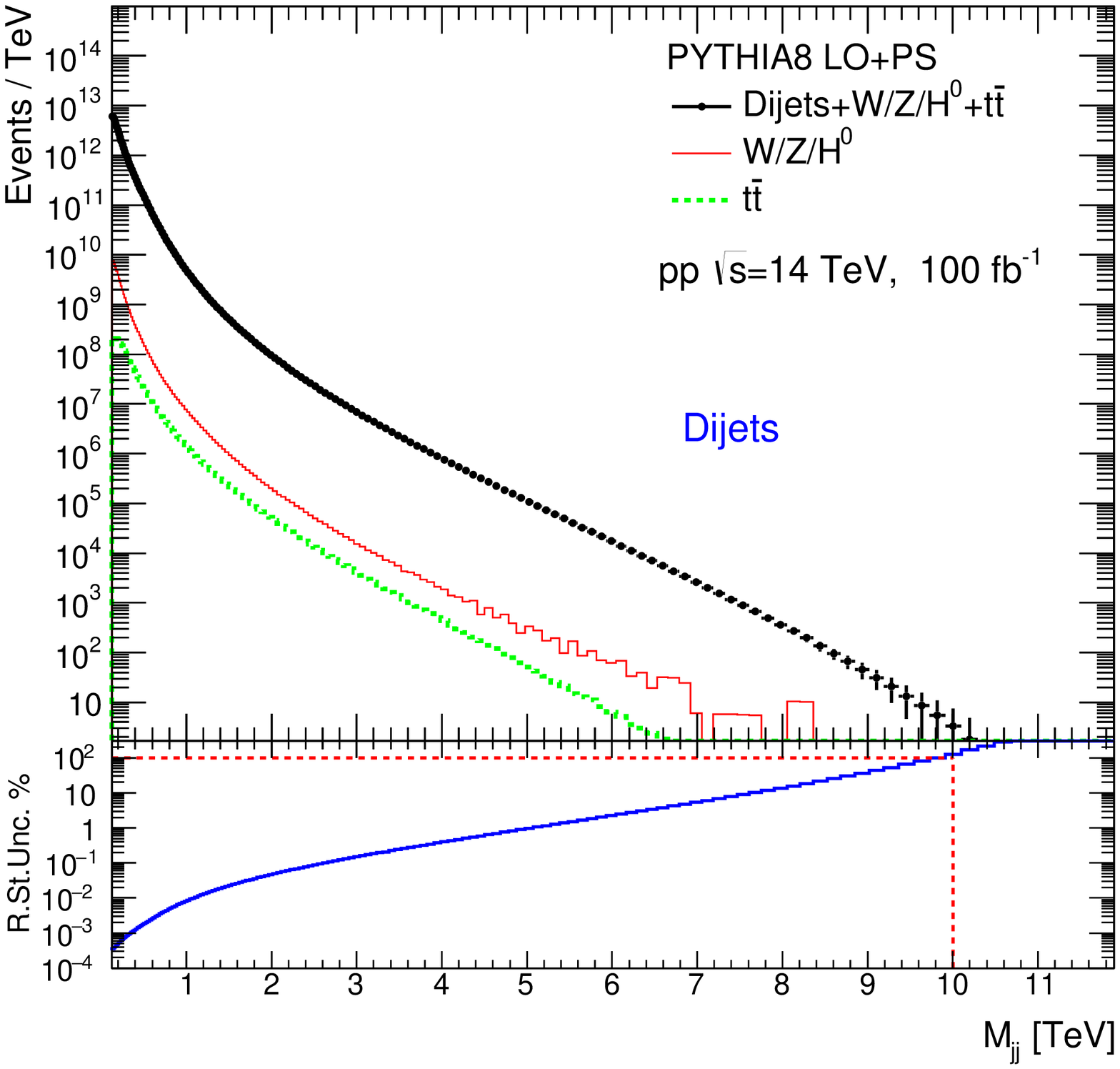}\hfill
   }
   \subfigure[ $3~\abb$] {
   \includegraphics[width=0.45\textwidth]{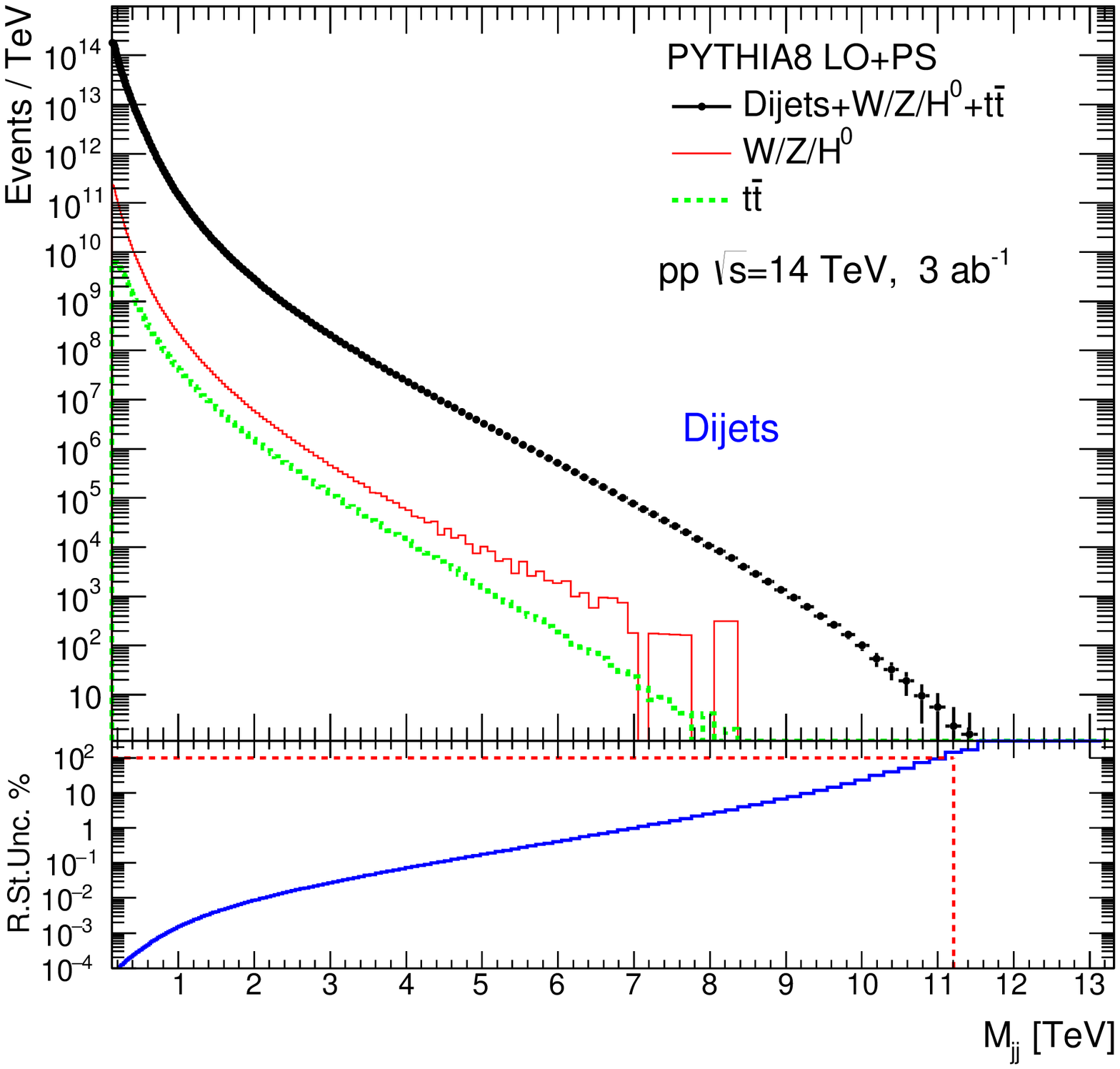}
   }
\end{center}
\caption{Expectations for the dijet invariant mass distribution for 100~$\fbb$ (3~$\abb$) for the 
LHC (HL-LHC) using the \pythia generator. Contributions
from  $W/Z/H^0$ -boson processes and top-quark processes are shown separately (without stacking the histograms).
The bottom plot shows the relative  statistical uncertainty for each bin, together with the line indicating the mass
point at which the uncertainty is $100\%$.  }
\label{fig:14tev_JetJetMass_2jet}
\end{figure}

Figure~\ref{fig:14tev_JetJetMass_2jet} shows the $\mjj$ distribution in \pythia for two
different integrated luminosities,  $100~\fbb$ and $3~\abb$.  
The bin size used in this figure 
gradually increases   
from 13~GeV for the lowest considered value of $\mjj$ to 190~GeV near $\mjj=10$~TeV. 
Such variable-size bins were previously used in \cite{Aad:2015xis,2016229, Aaboud:2017yvp} in order to  
minimize jet resolution effects,  and to reduce statistical
fluctuations in the tail of the $\mjj$  distribution. 

Figure~\ref{fig:14tev_JetJetMass_2jet} shows the sum of the three
contributions discussed in Sect.~\ref{sect_mc}, together with
the two contributions from $W/Z/H^0$-boson processes combined  and 
top-quark processes from the hard interactions (shown separately). 
The rate of the latter two processes combined near $\mjj=0.5$~TeV is 
$0.1\%$ of the total SM prediction.
At the same time, the contribution from the $t\bar{t}$ production is 
only $0.02\%$ of the total event rate.
The lower panel shows the relative statistical uncertainty on the 
data points, i.e. $\Delta d_i/d_i$, where $d_i$ is the number of 
the events in the bins, and $\Delta d_i$ its statistical uncertainty (which is $\sqrt{d_i}$ in the case of 
counting statistics). 

For a quantitative characterization of the dijet mass reach, 
we choose to define the $\mjj$ point at which the relative statistical uncertainty in a bin is 
100\% (or $\Delta d_i/d_i=1$), as indicated in the lower panel of 
Fig.~\ref{fig:14tev_JetJetMass_2jet} with the dash line.
In the case of counting statistics,  this  corresponds to one entry per bin.  
In this study, 
the point at  $\Delta d_i/d_i=1$  is  determined from many weighted events 
created by the \pythia event generator.

\begin{figure}
\begin{center}
   \subfigure[ No $y^*$ cut] {
   \includegraphics[width=0.45\textwidth]{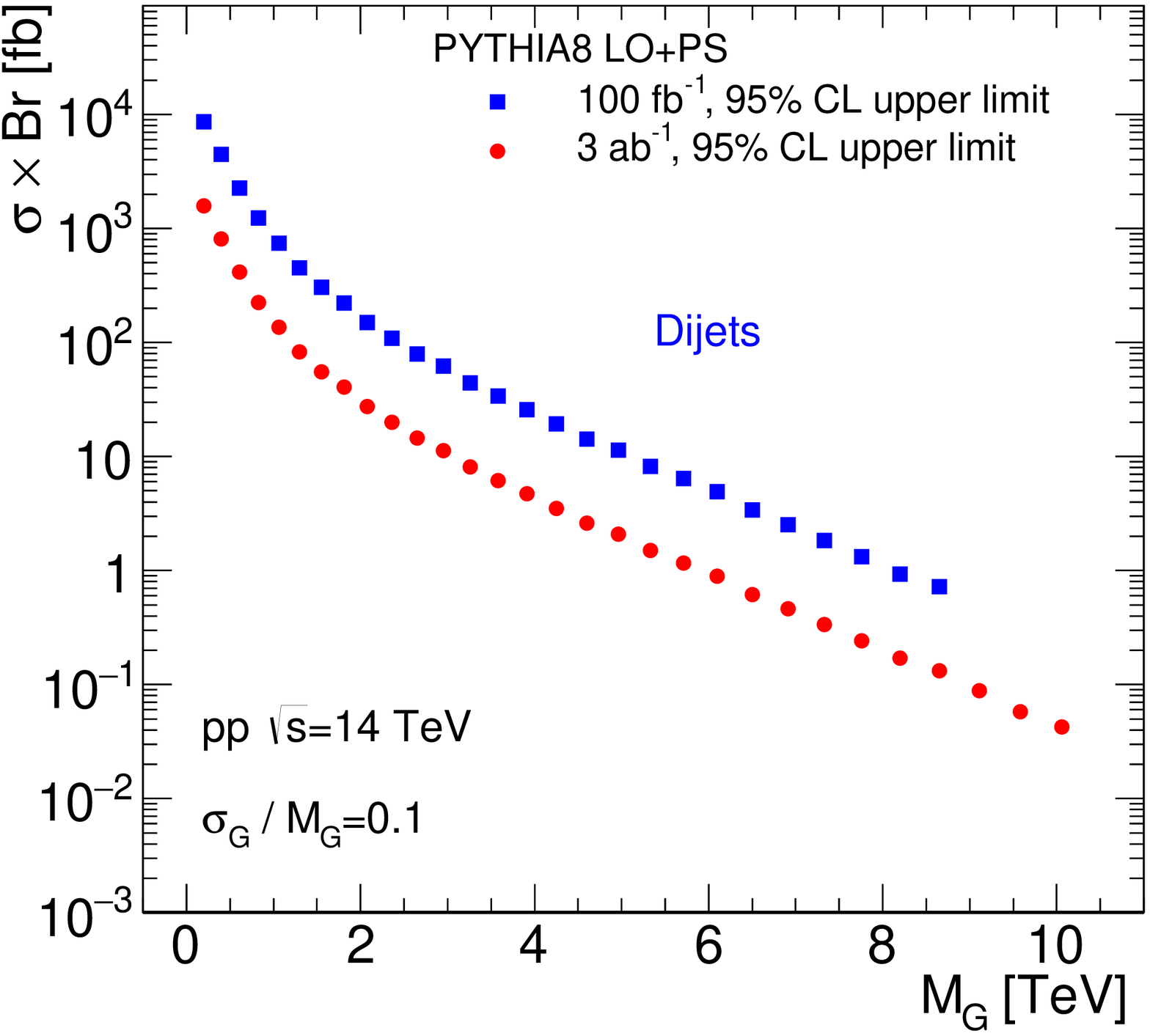}\hfill
   }
   \subfigure[ With $y^*$ cut] {
   \includegraphics[width=0.45\textwidth]{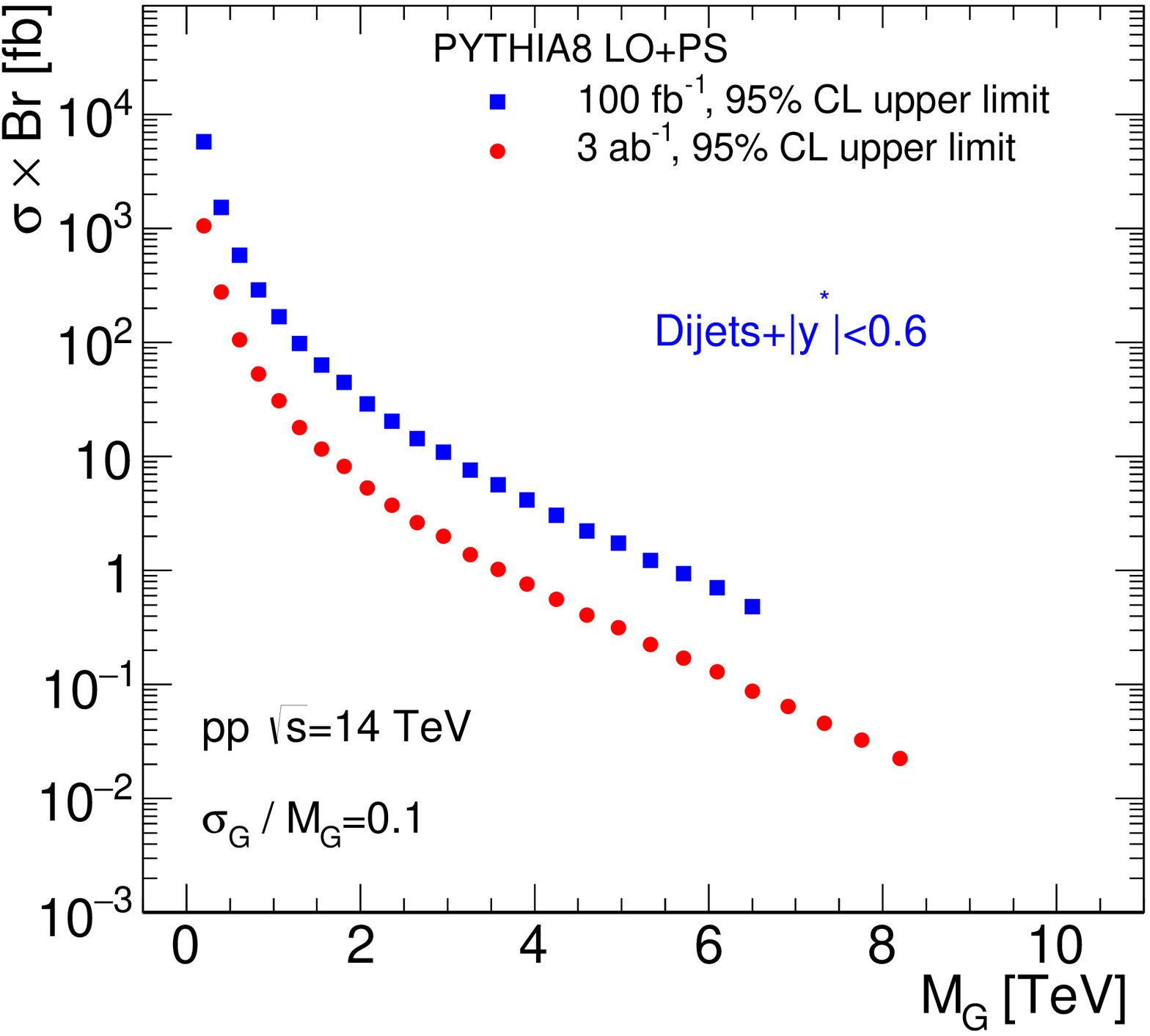}\hfill
   }
\end{center}
\caption{The 95\% C.L. upper limits obtained from the
$\mjj$  distribution on fiducial cross-section times the  branching ratio to two jets
for a hypothetical BSM signal approximated by a Gaussian contribution to the dijet mass spectrum.  The limits are obtained (a) without and (b) with $y^*$ cut.}
\label{fig:14tev_limits_JetJetMass_2jet}
\end{figure}

Figure~\ref{fig:14tev_limits_JetJetMass_2jet} shows the $95\%$ credibility-level (C.L.) upper limit
on fiducial cross-section times the  branching ratio for a generic Gaussian signal with
the width ($\sigma_G$) being  10\% of the Gaussian peak position.
The $95\%$ quantile of the posterior is taken as the upper limit on the possible 
number of signal events in data corresponding to that mass point. 
This value, divided by the corresponding luminosity, provides  
the upper limit on the production cross section of a new particle times the branching ratio (Br) to two jets.  
In addition to the inclusive jet case, we also calculate the upper limits after applying the rapidity difference requirement 
$|y^*|<0.6$ between two  jets \cite{ATLAS:2015nsi}  in order to enhance the sensitivity to heavy 
BSM particles decaying to jets. 

\begin{figure}
\begin{center}
 \includegraphics[width=0.45\textwidth]{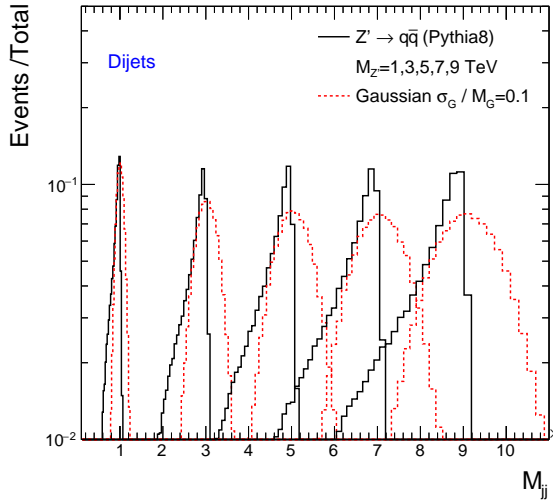}\hfill
\end{center}
\caption{The shapes of the $\mjj$ distributions for 
a hypothetical BSM signal approximated by a Gaussian contribution (red dash lines) and  
for the $Z'\to jets$  used for the calculation of the 95\% C.L. upper limits (black solid line).}
\label{fig:mjj_zprime}
\end{figure}

In order to calculate the expected upper limits for realistic shapes of the 
dijet mass distribution from heavy exotic particles,
such as $Z'$, we have performed a simulation of $Z'$ decays to jets, assuming the width of $15\%$ of the $Z'$ mass.  
The comparison of the Gaussian shape with the signal shape from the $Z'$ decays in 
\pythia is shown in Fig.~\ref{fig:mjj_zprime}.
The generated masses of the $Z'$ particles are given by the Breit-Wigner distribution,
but the $\mjj$ distributions are asymmetric, which are
typical for reconstructed  dijet masses using realistic jet algorithms (note the logarithmic scale used for this figure).

\begin{figure}
\begin{center}
   \subfigure[ $100~\fbb$] {
   \includegraphics[width=0.45\textwidth]{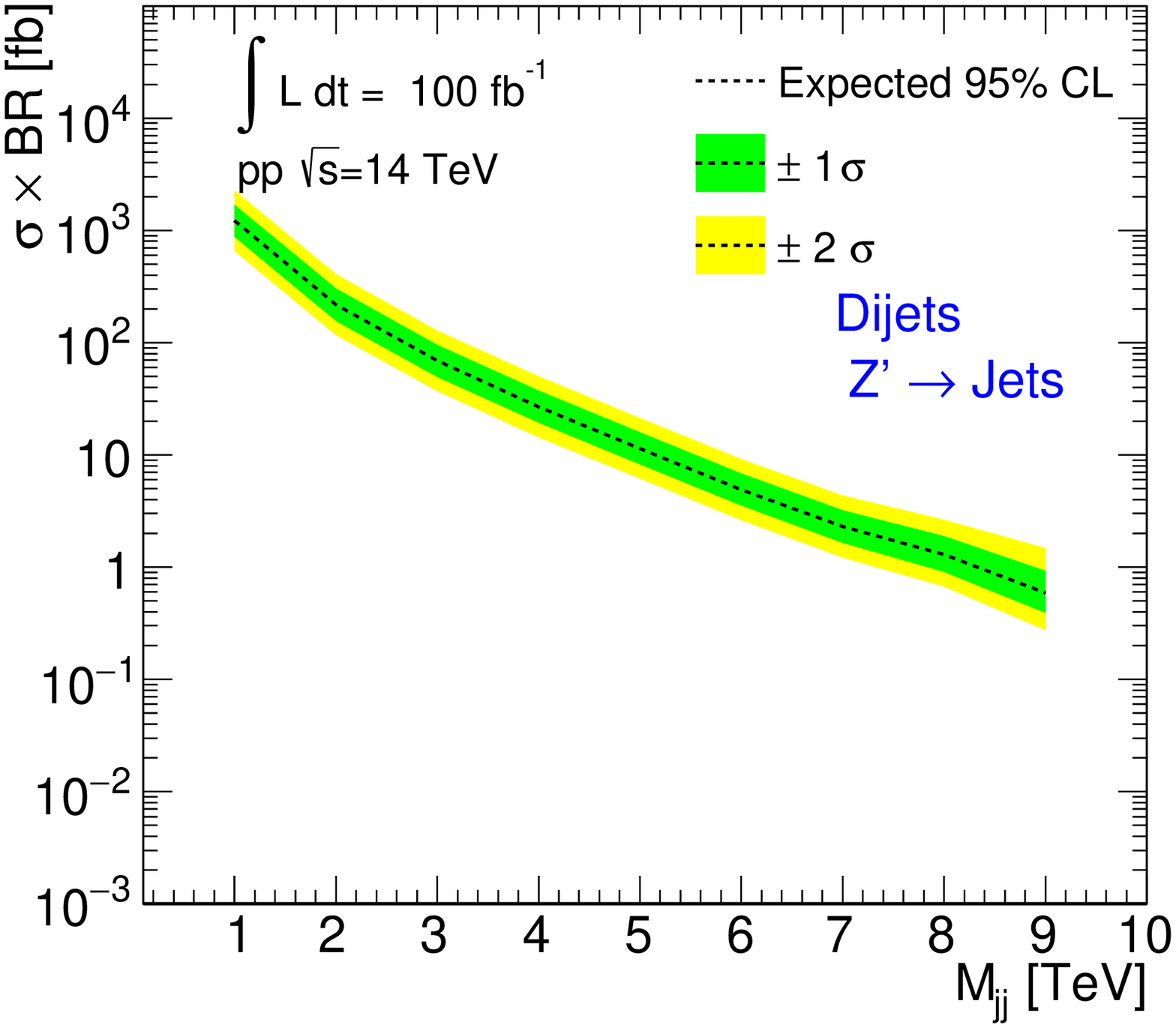}\hfill
   }
   \subfigure[ $3~\abb$] {
   \includegraphics[width=0.45\textwidth]{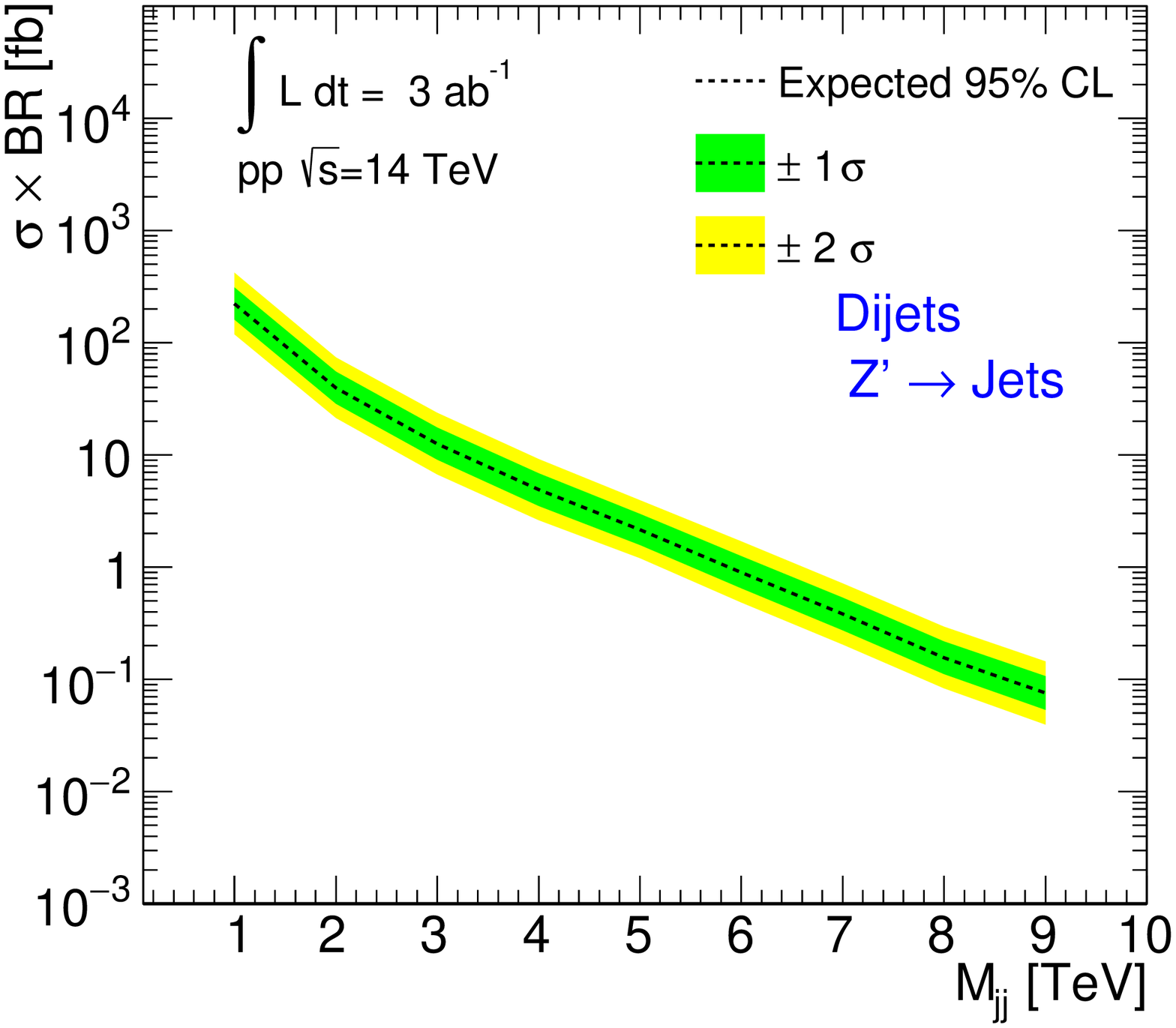}
   }
\end{center}
\caption{The 95\% C.L. upper limits obtained from the
$\mjj$  distribution on cross section times the  branching ratio to two jets
for a $Z'$ particle decaying to two jets.}
\label{fig:zprime_limits_JetJetMass_2jet}
\end{figure}

Alternatively, exclusion limits were calculated using the $CL_{s}$ method 
with a binned profile likelihood ratio as the test statistic using the {\sc HistFitter}  
framework \cite{Baak:2014wma}. The expected limits on the signal model are calculated by
using an asymptotic approximation \cite{Cowan:2010js}.
Figure~\ref{fig:zprime_limits_JetJetMass_2jet} shows the  95\% C.L. upper limits.
The green and yellow bands represent the $1 \sigma$ and $2 \sigma$ probability intervals around the expected limit.
The obtained limits are found to be rather similar to the Gaussian limits (without the requirement $|y^*|<0.6$) obtained above, 
despite the difference in the signal shape. Unlike the Gaussian limits,  the {\sc HistFitter}  limits were calculated starting 
from $\mjj>1$~TeV. Below this value, the {\sc HistFitter} technique  produces an unstable result  
caused by fluctuations in the $\mjj$ distribution after the extrapolation of the low-statistics histogram 
to the required luminosity (see Sect.~\ref{sect_mc}).

The limits shown in Fig.~\ref{fig:zprime_limits_JetJetMass_2jet} can be 
used for exclusion of models predicting peaks in the $\mjj$ distributions.   
Several BSM benchmark models, such as  
models of quantum black holes, excited quarks, $W'$ and $Z'$, have been  
excluded by CMS and ATLAS using 
LHC run I and run II data \cite{Aad:2015xis,2016229, Aaboud:2017yvp, Khachatryan:2010jd, Chatrchyan:2013qha, Sirunyan:2016iap}. 
Therefore, we do not show the cross sections for such BSM  models 
in Fig.~\ref{fig:14tev_limits_JetJetMass_2jet} and \ref{fig:zprime_limits_JetJetMass_2jet}.

\begin{figure}
\begin{center}
   \subfigure[ $100~\fbb$] {
   \includegraphics[width=0.45\textwidth]{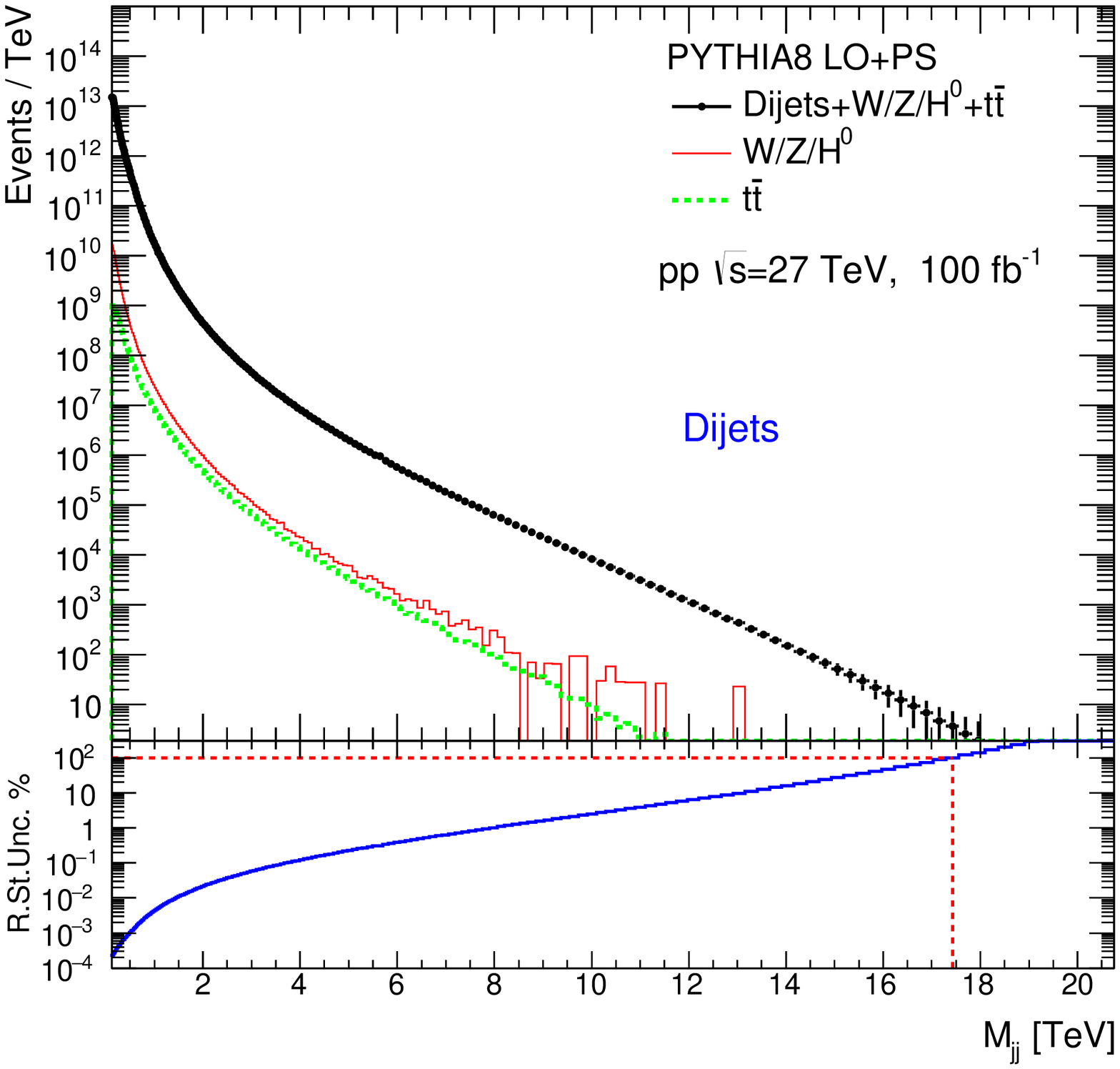}\hfill
   }
   \subfigure[ $3~\abb$] {
   \includegraphics[width=0.45\textwidth]{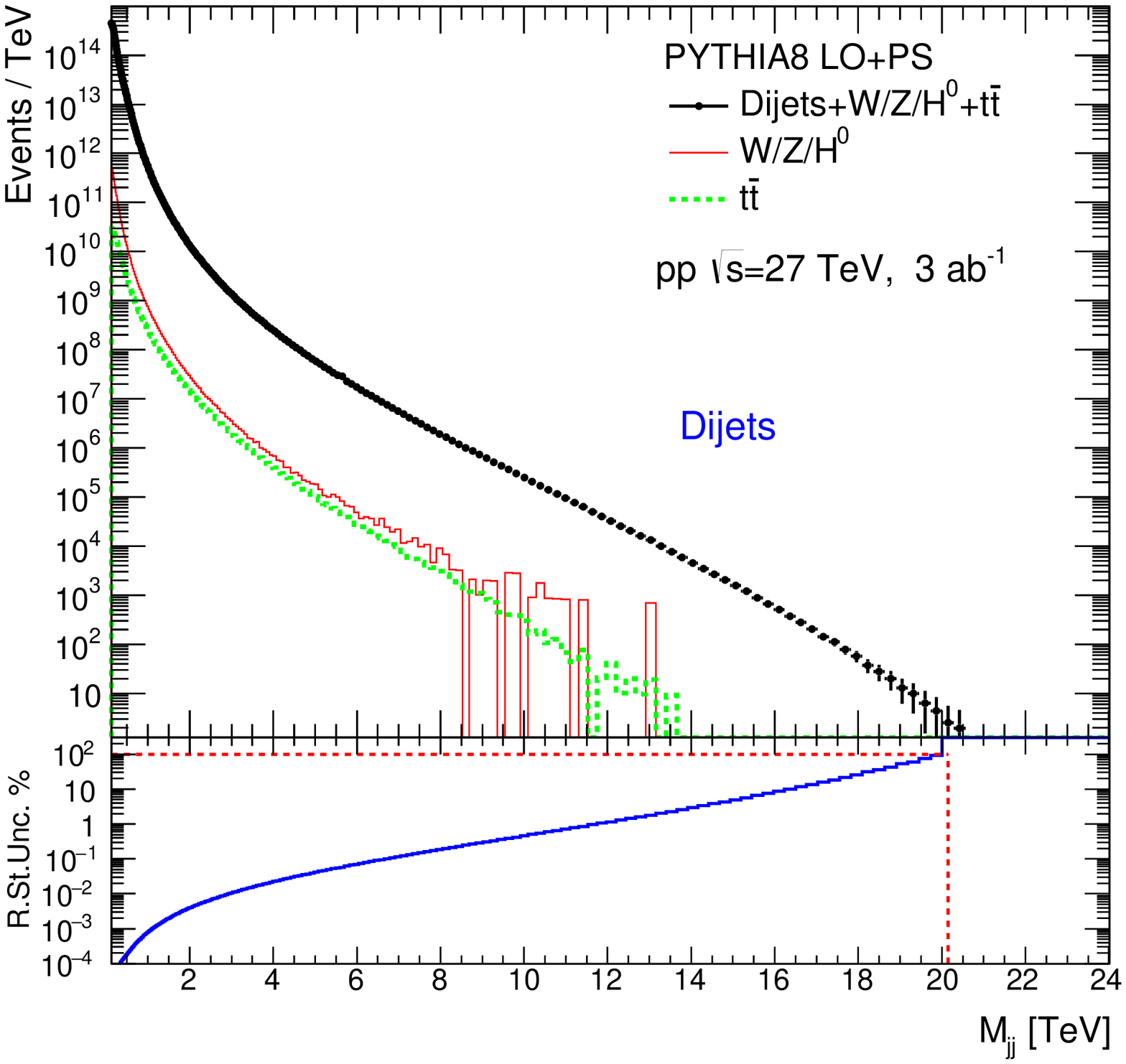}
   }
\end{center}
\caption{The distribution of the dijet invariant masses  for 100~$\fbb$ and 3~$\abb$ at the HE-LHC, together with the relative statistical uncertainty shown in bottom panel.}
\label{fig:27tev_JetJetMass_2jet}
\end{figure}

\begin{figure}
\begin{center}
   \includegraphics[width=0.45\textwidth]{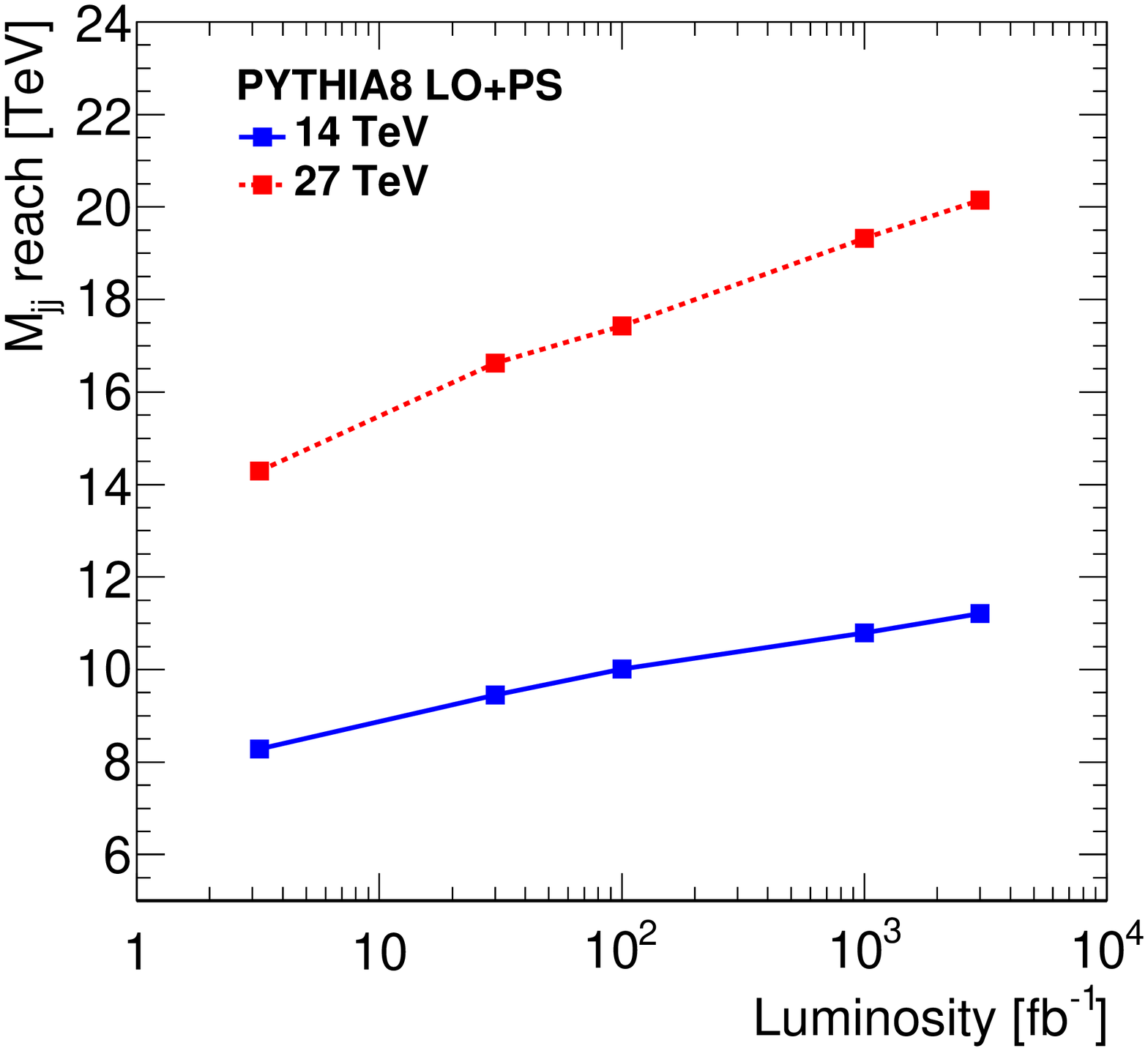}\hfill
\end{center}
\caption{Dijet mass reach for the HL-LHC and HE-LHC experiments. The uncertainties on the data 
points, derived from the bin width used for the simulated $\mjj$ distributions, are compatible with the size of the symbols.
}
\label{fig:mass_reach}
\end{figure}

The \pythia expectations  for the $\mjj$ distribution for the  HE-LHC 
are  shown in Fig.~\ref{fig:27tev_JetJetMass_2jet}.
The dijet mass distribution uses bin sizes  
that gradually increase
from 13~GeV for the lowest value of $\mjj$ to 280~GeV near $\mjj=10$~TeV.
The lower panel shows the relative statistical uncertainties together with the line 
indicating the mass at which the relative statistical uncertainty on the data point is $100\%$. 
Figure~\ref{fig:mass_reach} shows the mass reach as a function of integrated luminosity for the 
HL-LHC and HE-LHC, defined by the point at  which the  relative statistical
uncertainty is  $100\%$. 
The $\mjj$ mass reach at the centre-of-mass of 27~TeV is close to 
$17$~TeV,  even for the modest luminosity of $100~\fbb$. This is a factor of two 
larger than the dijet mass reach for the luminosity expected at the HL-LHC.

\begin{figure}
\begin{center}
   \subfigure[ No $y^*$ cut] {
   \includegraphics[width=0.45\textwidth]{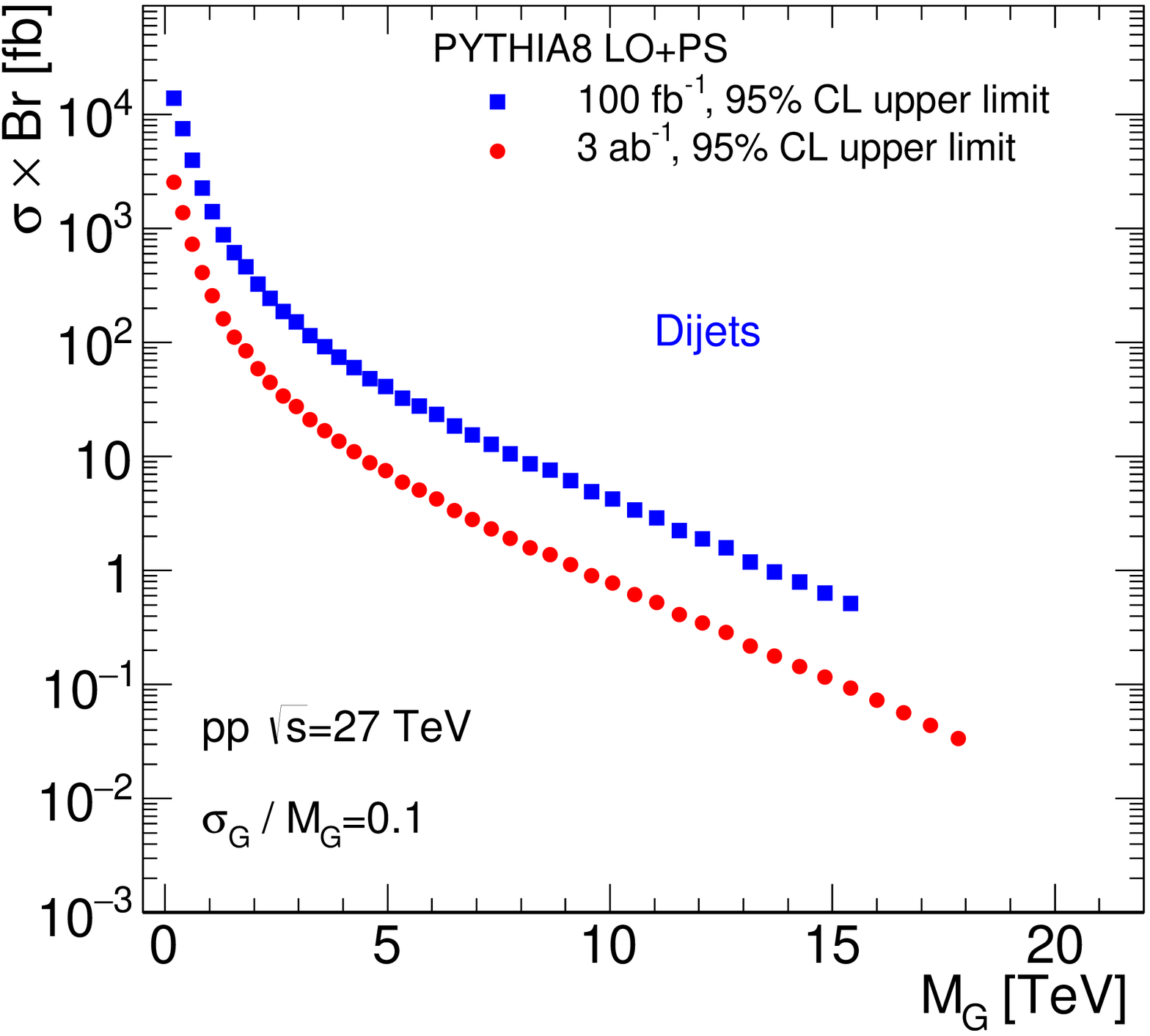}\hfill
   }
   \subfigure[ With $y^*$ cut] {
   \includegraphics[width=0.45\textwidth]{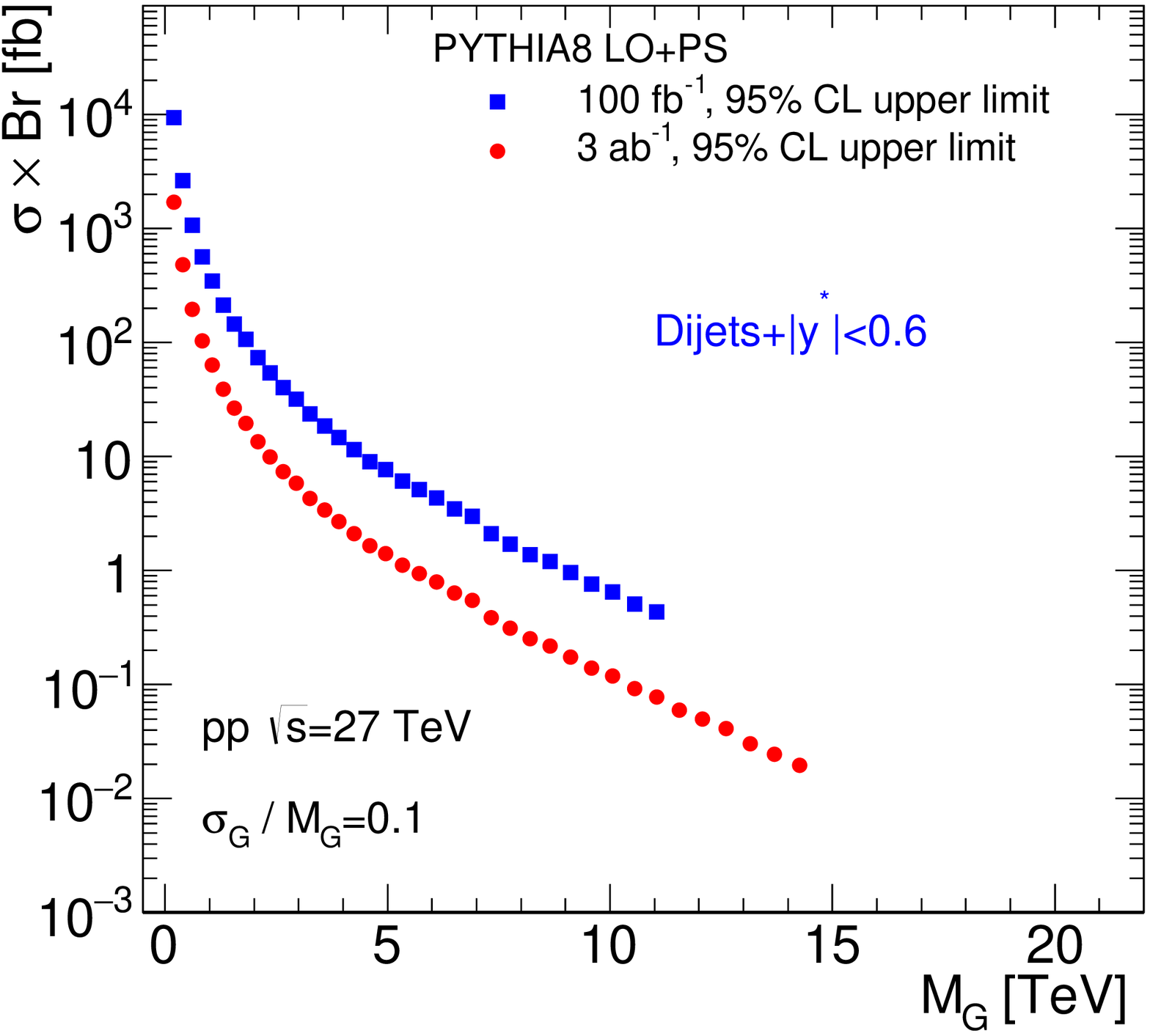}\hfill
   }
\end{center}
\caption{The 95\% C.L. upper limits obtained from the
$\mjj$  distribution on the cross section times the  branching ratio to two jets
for a hypothetical signal approximated by a Gaussian contribution to the expected dijet mass.
The HE-LHC expectations were obtained using the \pythia generator.
}
\label{fig:27tev_limits_JetJetMass_2jet}
\end{figure}

Figure~\ref{fig:27tev_limits_JetJetMass_2jet} shows the 95\% C.L.  upper limits
on  the product of the cross section and the branching ratio  
for a signal approximated by a Gaussian whose width is 10\% of the mass of the searched resonance. 
The expected limit for $3~\abb$ is a factor of ten better than for $100~\fbb$.  

%\input{sect_dijets_mu.tex}
%%%%%%%%%%%%%%%%%%%%%%%%%%%%%%%%%%%%%%%%%%%%%%%%%%%%%%%%%%%%%%%%%%
\section{Dijets in events with associated muons}
\label{sect_dijets_mu}
%%%%%%%%%%%%%%%%%%%%%%%%%%%%%%%%%%%%%%%%%%%%%%%%%%%%%%%%%%%%%%%%%%

The previous studies of inclusive dijets represent a hypothetical scenario
that may never be realized in practice due to difficulties \footnote{Data analyses that use
a complex trigger menu with a mixture of prescaled and unprescaled triggers are possible, but such studies require a significant effort and, as the result, are rather rare.}
in analyzing data with trigger prescales  applied
to jets at medium $p_T$. 
However, measurements of unbiased $\mjj$ distributions  (below 1~TeV) can be made possible by using
an independent object to trigger on.
For example, to reduce the impact of the very high rate of multi-jet background, 
at the price of requiring associated production, one can require an isolated muon, electron or other particle.

\begin{figure}[h] 
\begin{center}
   \subfigure[With truth-level leptons] {
   \includegraphics[width=0.45\textwidth]{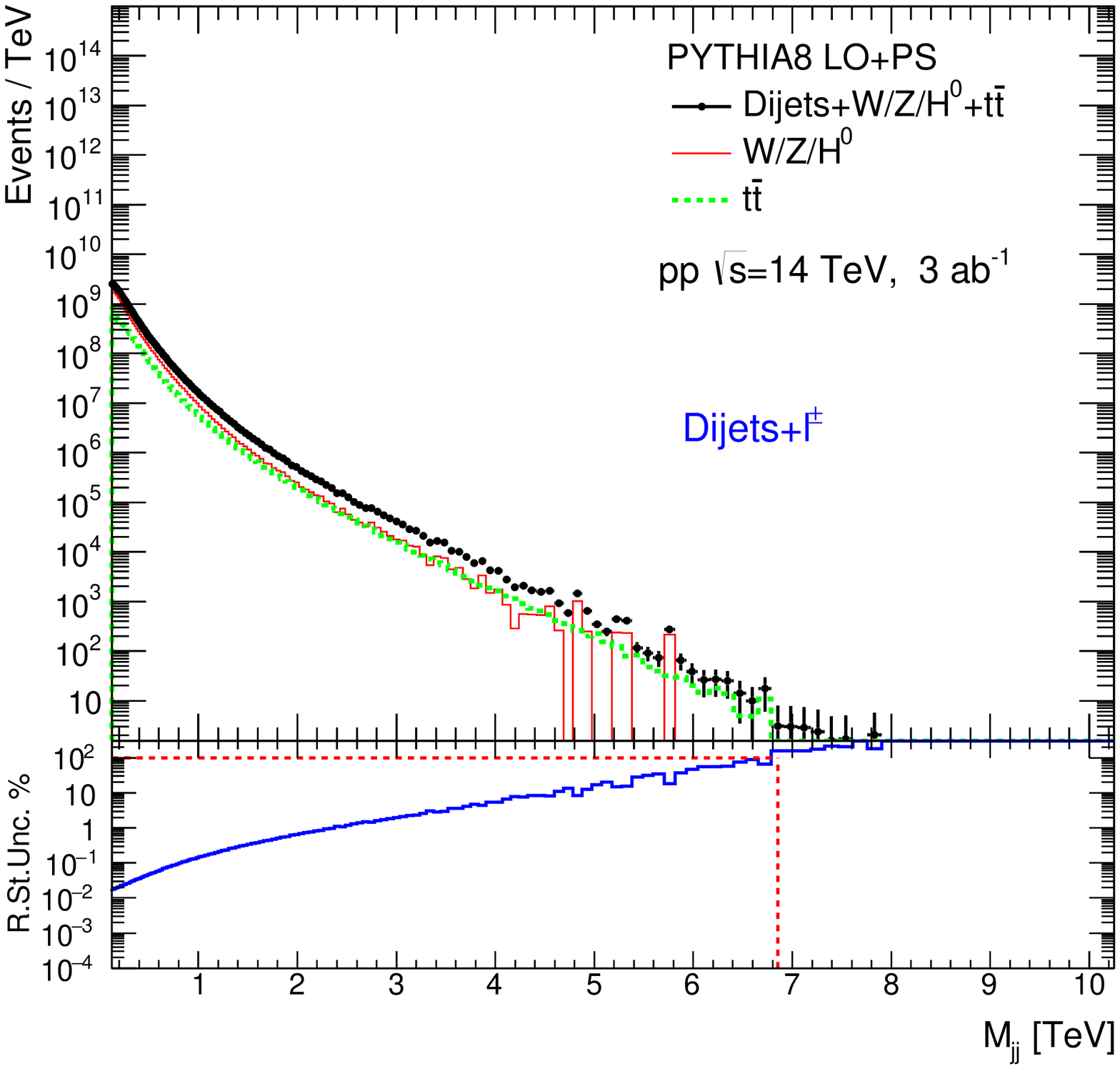}\hfill
   }
   \subfigure[With truth-level muons plus mis-identified jets ] {
   \includegraphics[width=0.45\textwidth]{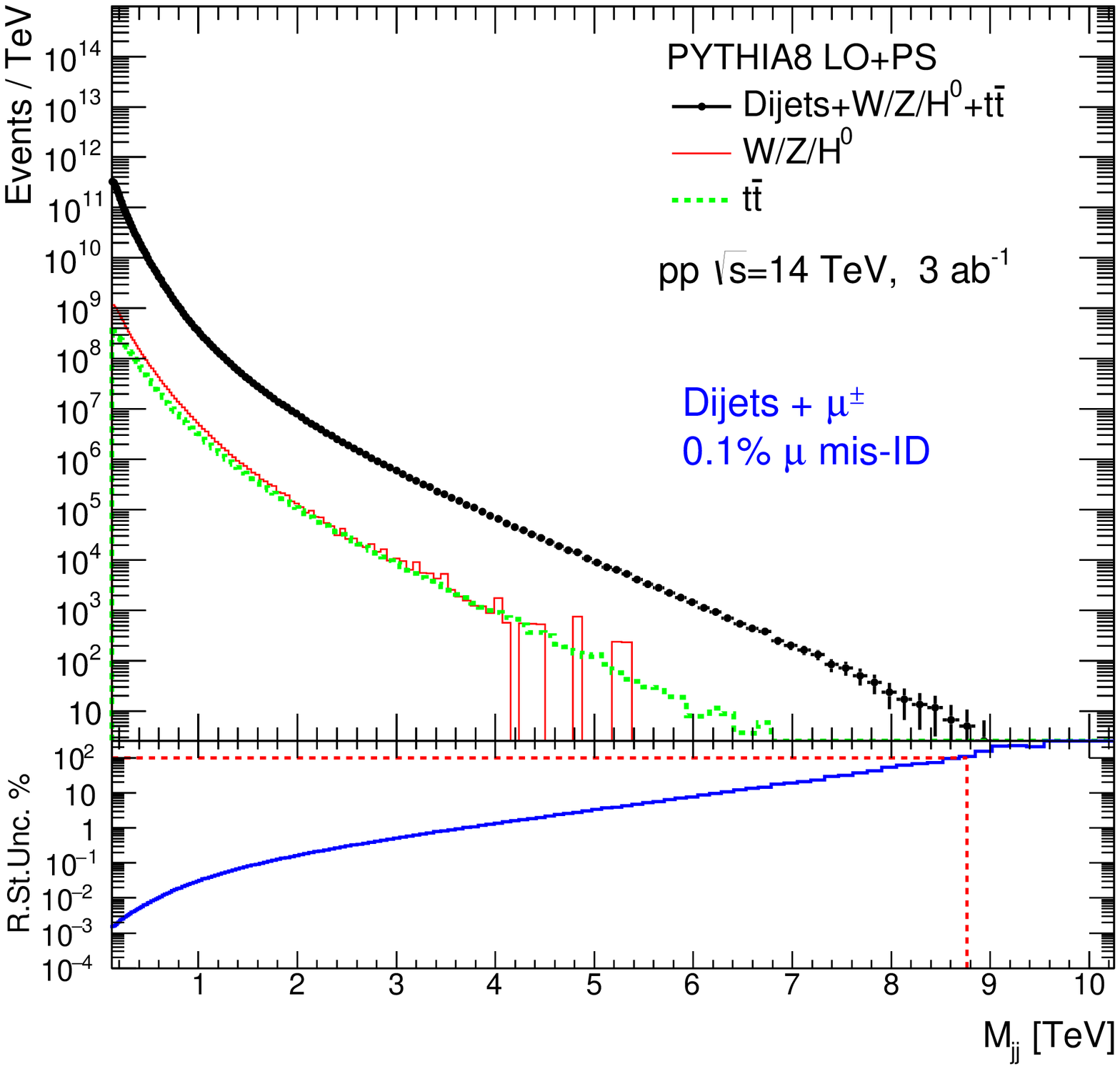}
   }
\end{center}
\caption{Dijet invariant masses with associated leptons for 3~$\abb$ of the HL-LHC experiment.
The distributions are shown for (a) dijets with muons and electrons without mis-identification and (b)
for muons assuming  0.1\% mis-identification rate.}
\label{fig:14tev_JetJetMass_2jet_mu}
\end{figure}

Figure~\ref{fig:14tev_JetJetMass_2jet_mu}(a) shows the dijet invariant masses with 
associated leptons with  $p_T(l)>60$~GeV using  the isolation requirements as
described in  Sect.~\ref{sec:reco}. 
This figure corresponds to the ideal case when the lepton mis-identification rate is set to  zero.
The fraction of  the combined $W/Z/H^0$ and top events to the total predicted event rate  is $96\%$.
This shows that any new physics that leads to resonances with the production rates 
compatible with $W/Z/H^0$ processes can easily be detected,
unlike the case with fully inclusive jets discussed in the previous section. 

\begin{figure}[h] 
\begin{center}
   \subfigure[ $100~\fbb$] {
   \includegraphics[width=0.45\textwidth]{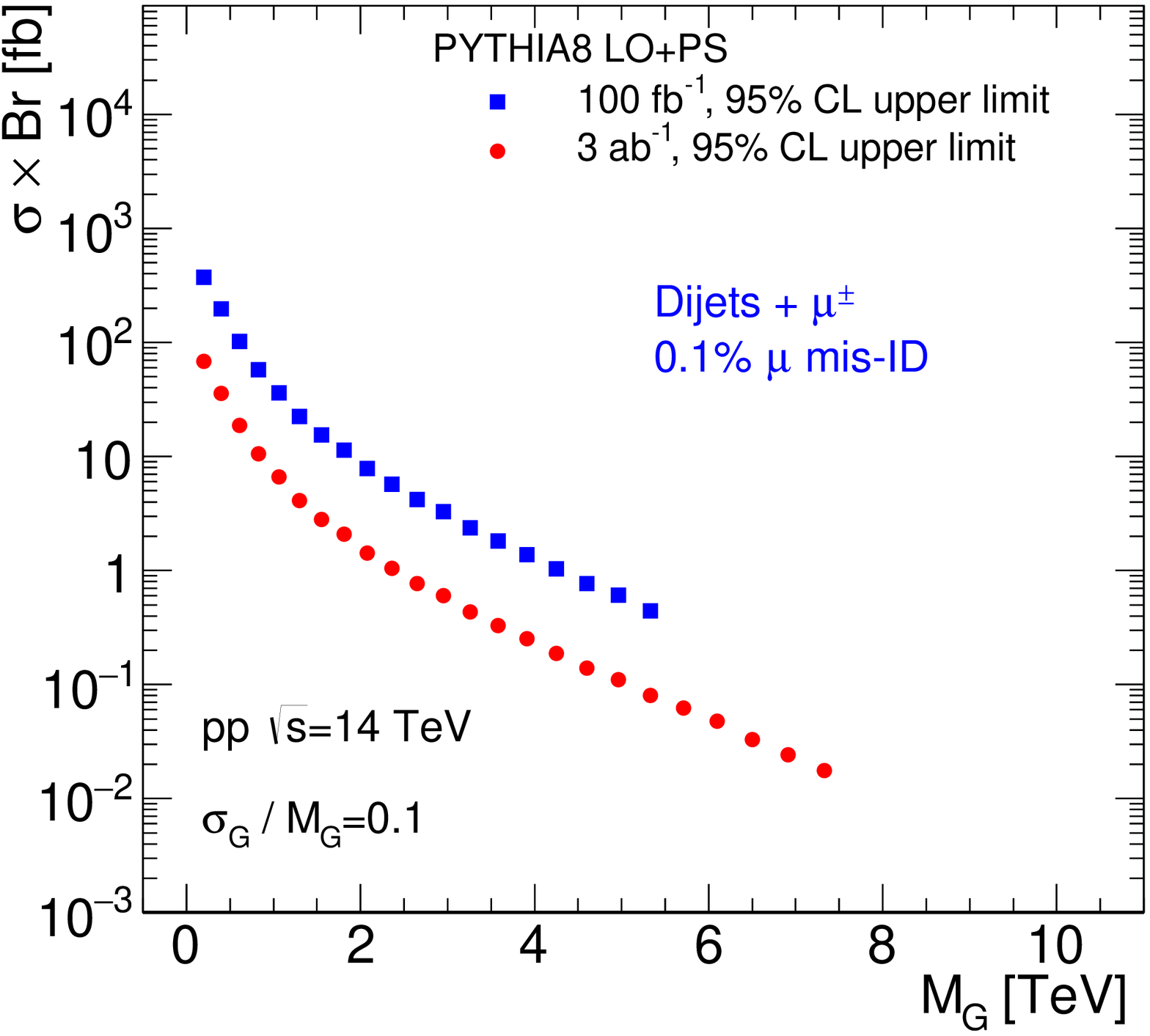}\hfill
   }
   \subfigure[ $3~\abb$] {
   \includegraphics[width=0.45\textwidth]{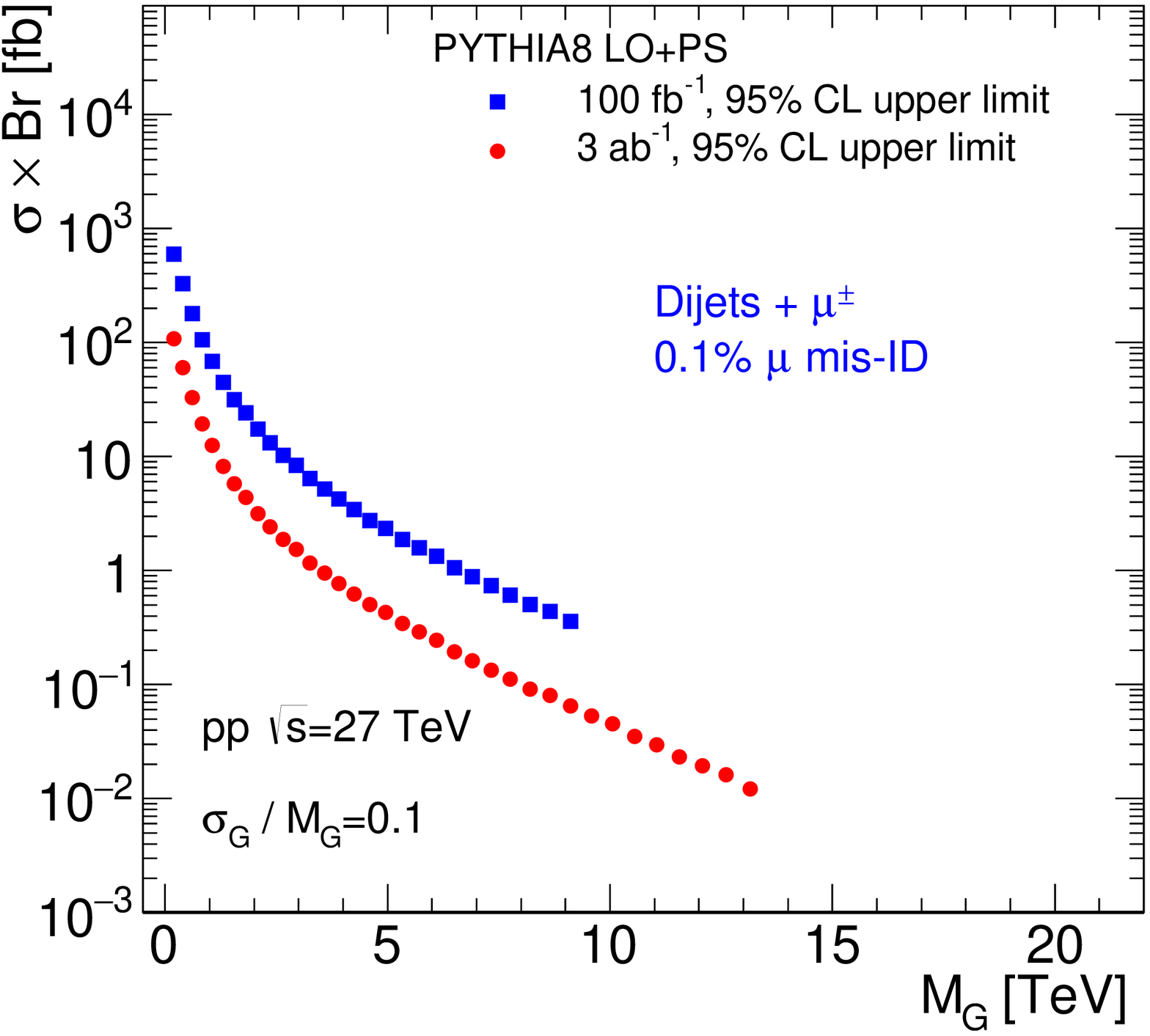}
   }
\end{center}
\caption{The 95\% C.L. upper limits obtained from the
$\mjj$  distribution on cross-section times the  branching ratio to two jets with isolated muons
for a hypothetical BSM signal approximated by a Gaussian contribution to the expected dijet mass. The limits were calculated
for the HL-LHC and HE-LHC.}
\label{fig:limits_JetJetMass_2jet_lep_mu}
\end{figure}

To illustrate a scenario with lepton mis-identifications, 
let us now turn to the case with muons.
Figure~\ref{fig:14tev_JetJetMass_2jet_mu}(b) shows 
the $\mjj$ distribution with isolated muons, including contributions from jets
which are mis-identified as muons. The mi-identification rate is set to $0.1\%$ as discussed in Sect.~\ref{sec:reco}.
According to Fig.~\ref{fig:14tev_JetJetMass_2jet_mu}(b), the fraction of $W/Z/H^0/$top processes is $1.2\%$ to 
the total event rate, which is  
a factor of ten larger than for the inclusive dijets shown in Fig.~\ref{fig:14tev_JetJetMass_2jet}.
Figure~\ref{fig:14tev_JetJetMass_2jet_mu}(a) and (b) shows that
the contribution from 
EWK and top processes to the total event rate strongly depends 
on the mis-identification rates for leptons. For realistic scenarios, 
exact fractions of EWK and top contributions should be calculated using full detector simulations. 

\begin{figure}[h] 
\begin{center}
   \subfigure[ $100~\fbb$] {
   \includegraphics[width=0.45\textwidth]{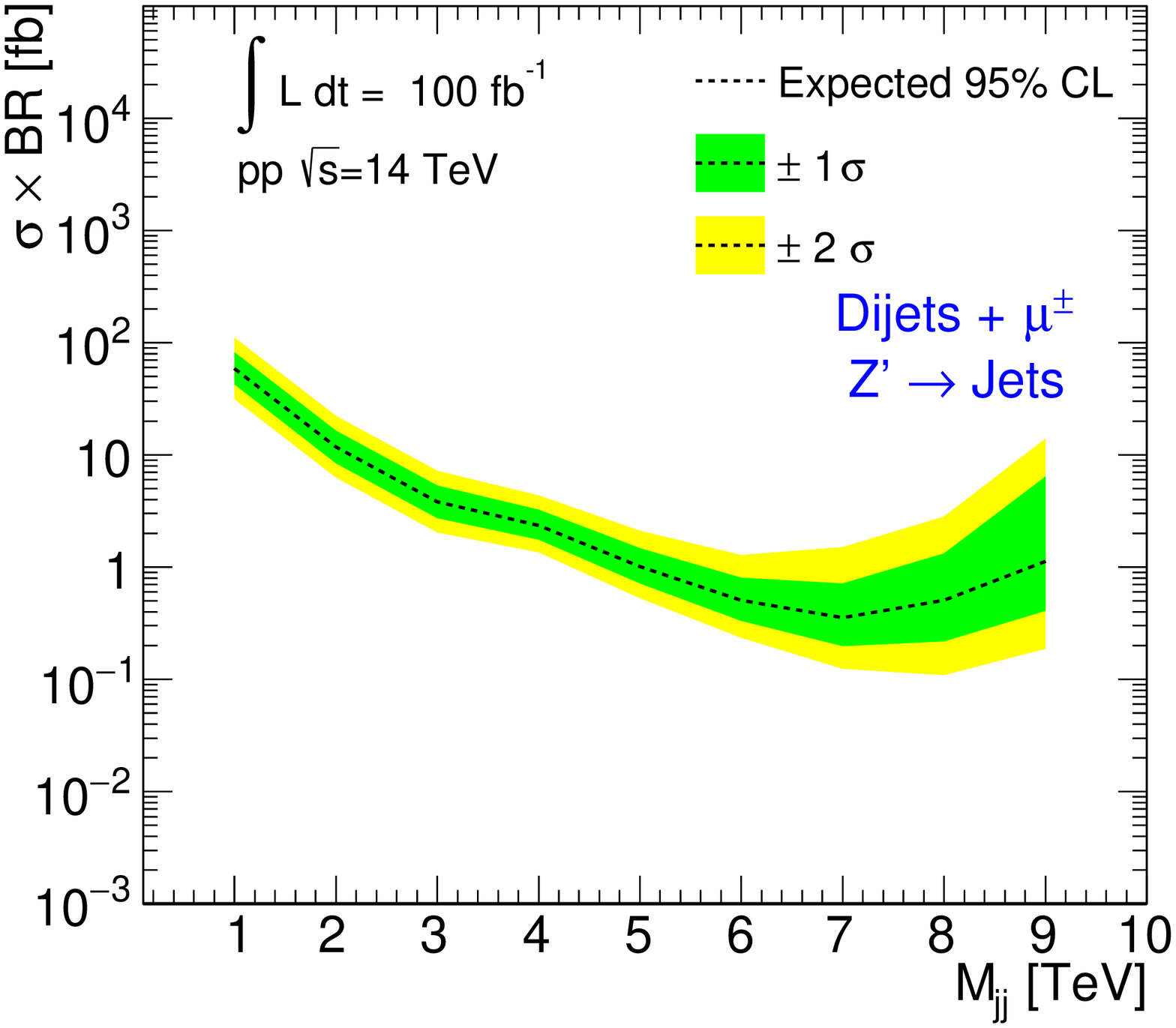}\hfill
   }
   \subfigure[ $3~\abb$] {
   \includegraphics[width=0.45\textwidth]{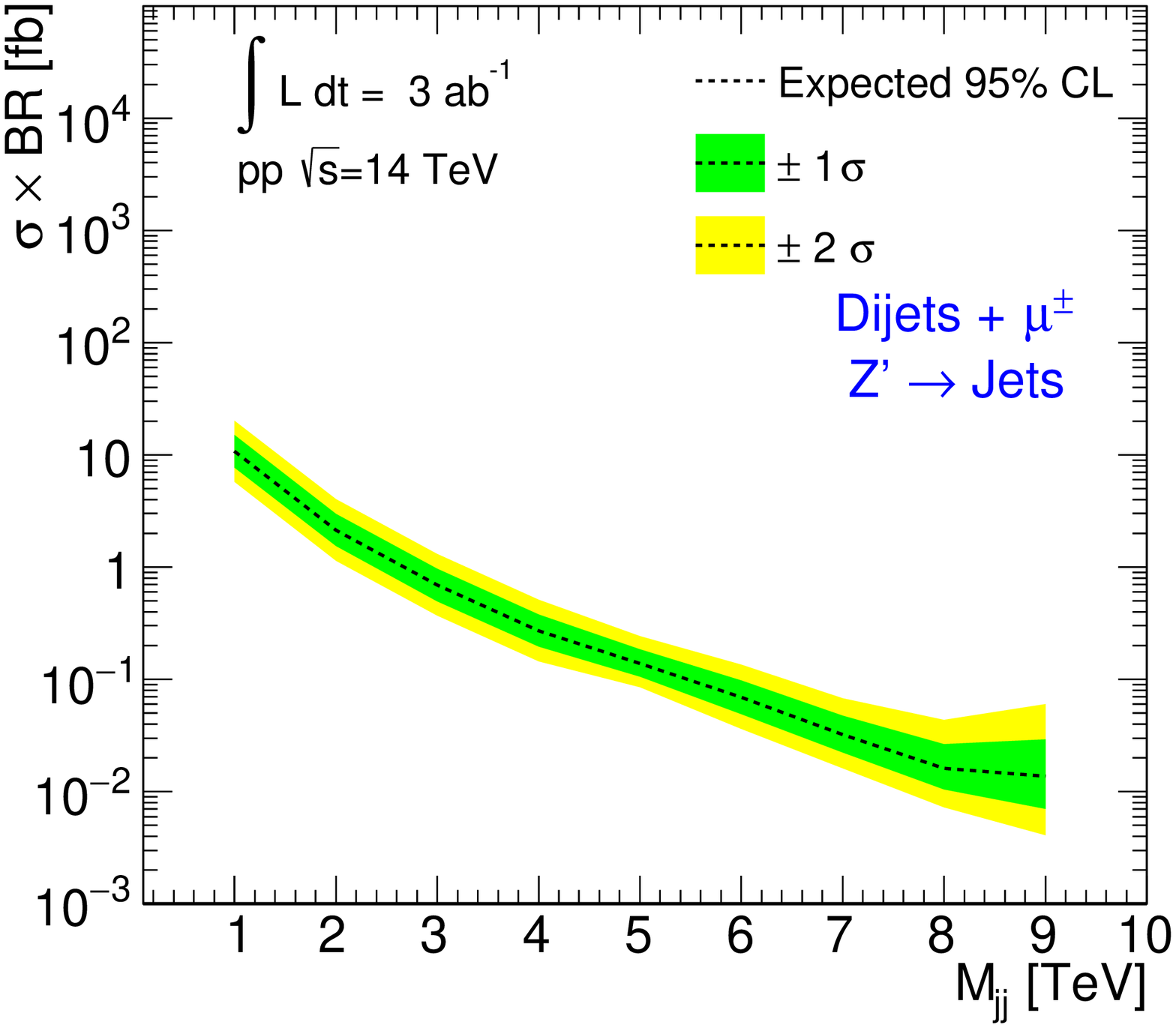}
   }
\end{center}
\caption{
The 95\% C.L. upper limits obtained from the
$\mjj$ distribution on the cross-section times branching ratio for a hypothetical BSM process producing 
a resonance in the dijet mass distribution and is produced in association with an isolated  muon, 
approximating the width of the dijet resonance by a Gaussian distribution.} 
\label{fig:zprime_limits_JetJetMass_2jet_mu}
\end{figure}

As for  the previous sections, our goal is to give expectations for exclusion
limits at the HL-LHC and HE-LHC using assumed muon fake rates, without discussing the cross sections for
particular exotic models
predicting enhancements in dijet masses.
The 95\% credibility-level upper limits for 
a signal for the muon-associated dijet production, 
assuming $0.1\%$ mis-identification rate, are shown in Fig.~\ref{fig:limits_JetJetMass_2jet_lep_mu}. 
The signal was 
approximated by a Gaussian distribution whose width is 10\% of the mass of the searched resonance. 
The figure shows the expectations for the HL-LHC and HE-LHC. 
Figure~\ref{fig:zprime_limits_JetJetMass_2jet_mu} shows the  95\% credibility-level upper limits for 
a $Z'$ signal comparing $100~\fbb$ from the LHC with the HL-LHC luminosity scenario.

%\input{sect_bjets.tex}
%%%%%%%%%%%%%%%%%%%%%%%%%%%%%%%%%%%%%%%%%%%%%%%%%%%%%%%%%%%%%%%%%%
\section{Studies of $b$-jets at the HL-LHC and HE-LHC}
\label{sect_bjets}
%%%%%%%%%%%%%%%%%%%%%%%%%%%%%%%%%%%%%%%%%%%%%%%%%%%%%%%%%%%%%%%%%%

\begin{figure}[h]
\begin{center}
   \subfigure[ $100~\fbb$] {
   \includegraphics[width=0.45\textwidth]{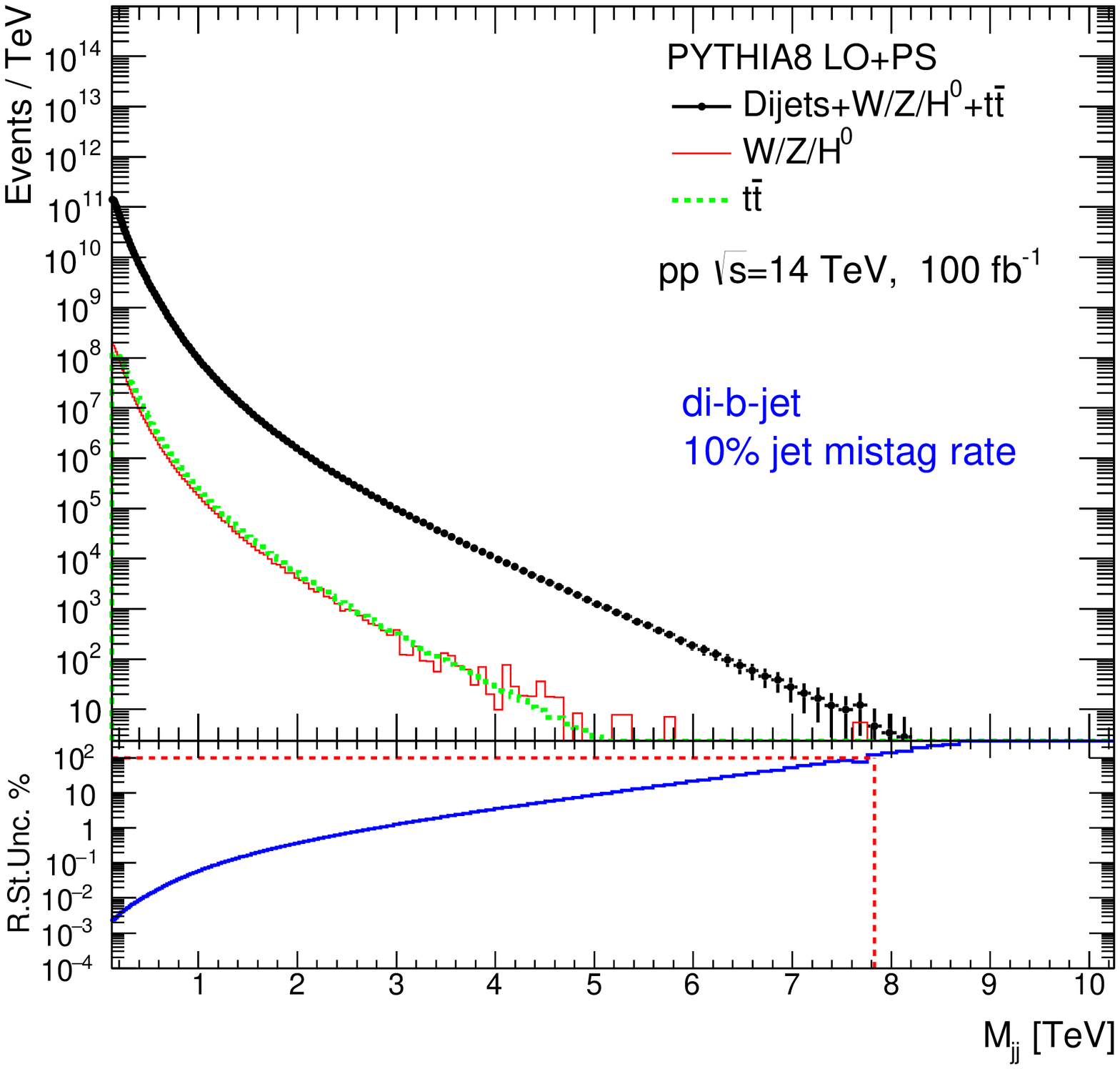}\hfill
   }
   \subfigure[ $3~\abb$] {
   \includegraphics[width=0.45\textwidth]{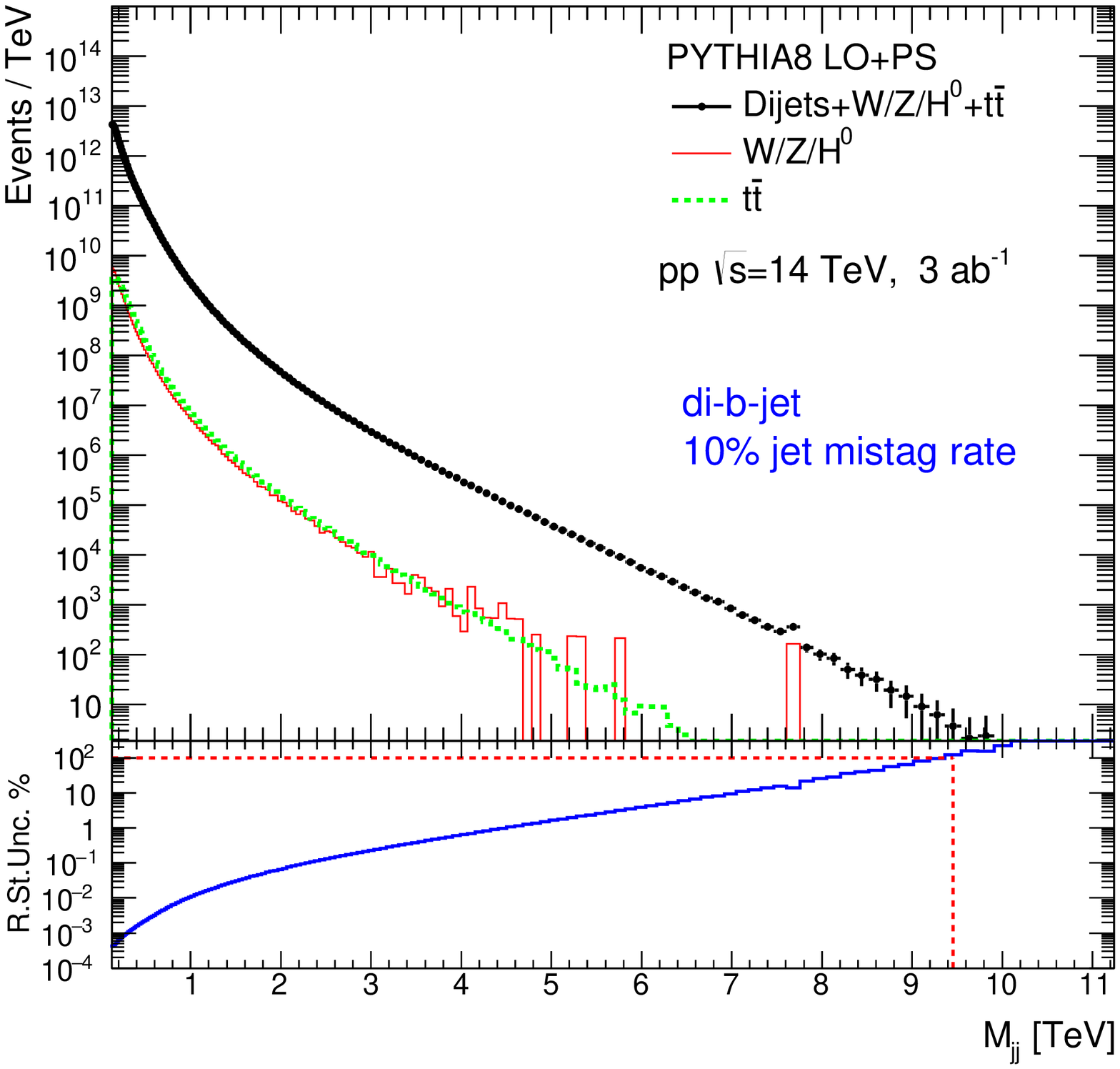}
   }
\end{center}
\caption{The $\mjj$ distributions for jets identified as $b$-jets, 
for 100~$\fbb$ and 3~$\abb$, together with the relative statistical uncertainty shown in bottom panel.}
\label{fig:14tev_JetJetMass_2bjet}
\end{figure}

Now we consider the case with $b-$jets selected as described in Sec.~\ref{sec:reco}.
Figure~\ref{fig:14tev_JetJetMass_2bjet} shows the $\mjj$  predictions for the sum of the three
contributions (QCD dijets, vector/scalar boson and top production) discussed in Sect.~\ref{sect_mc}, together with
the two contributions from $W/Z/H^0$ -boson processes  and top-quark processes from the hard interactions. The total event rate is about 2\% of the inclusive dijet after $b$-tagging.
The rate of the $W/Z/H^0$ and top processes combined near $\mjj=0.5$~TeV is about $0.2\%$ of  the total event rate.
The contribution from the $t\bar{t}$ production is larger than that from the $W/Z/H^0$-boson processes.

\begin{figure}[h]
\begin{center}
   \subfigure[ No $y^*$ cut] {
   \includegraphics[width=0.45\textwidth]{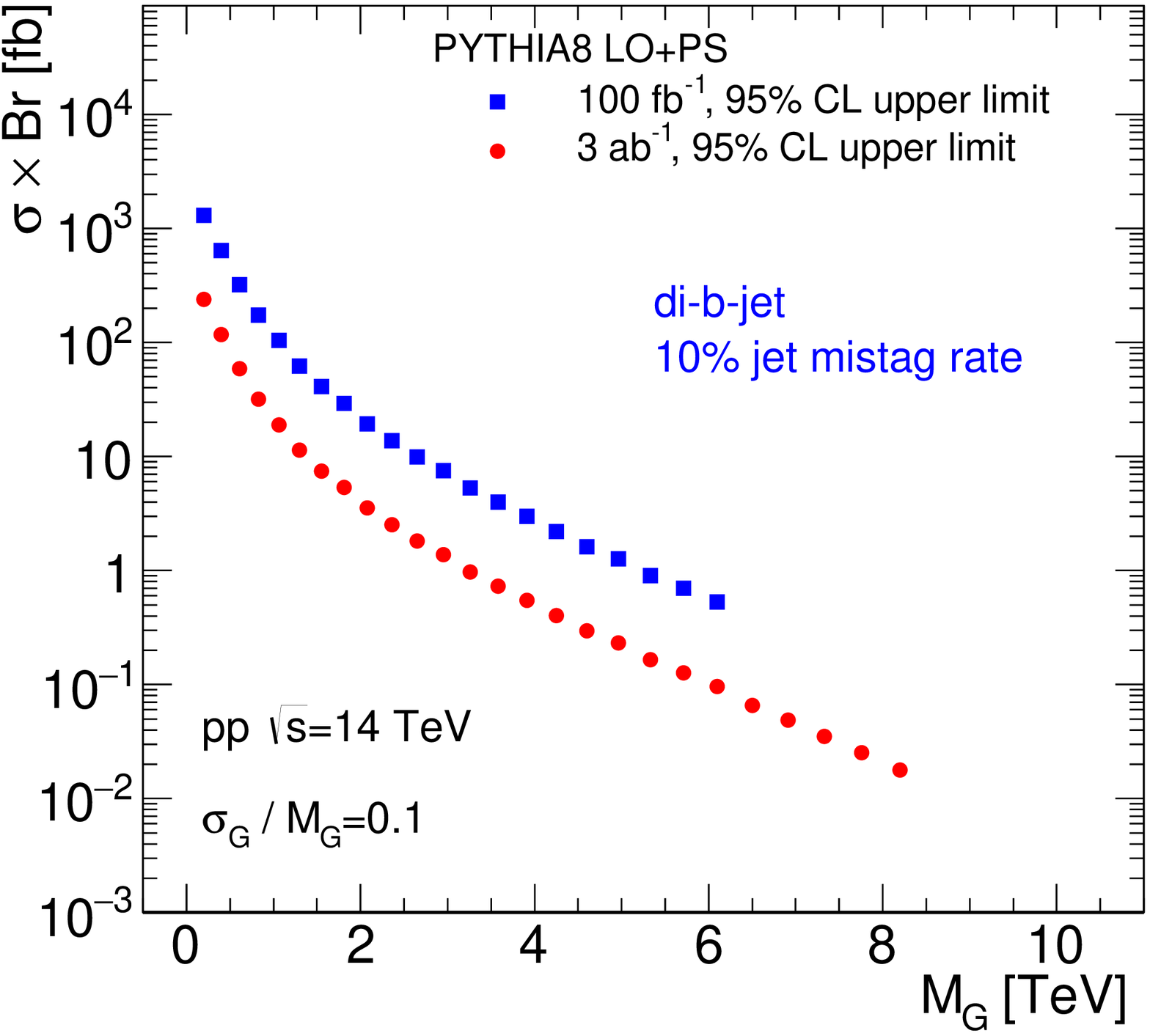}\hfill
   }
   \subfigure[ With $y^*$ cut] {
   \includegraphics[width=0.45\textwidth]{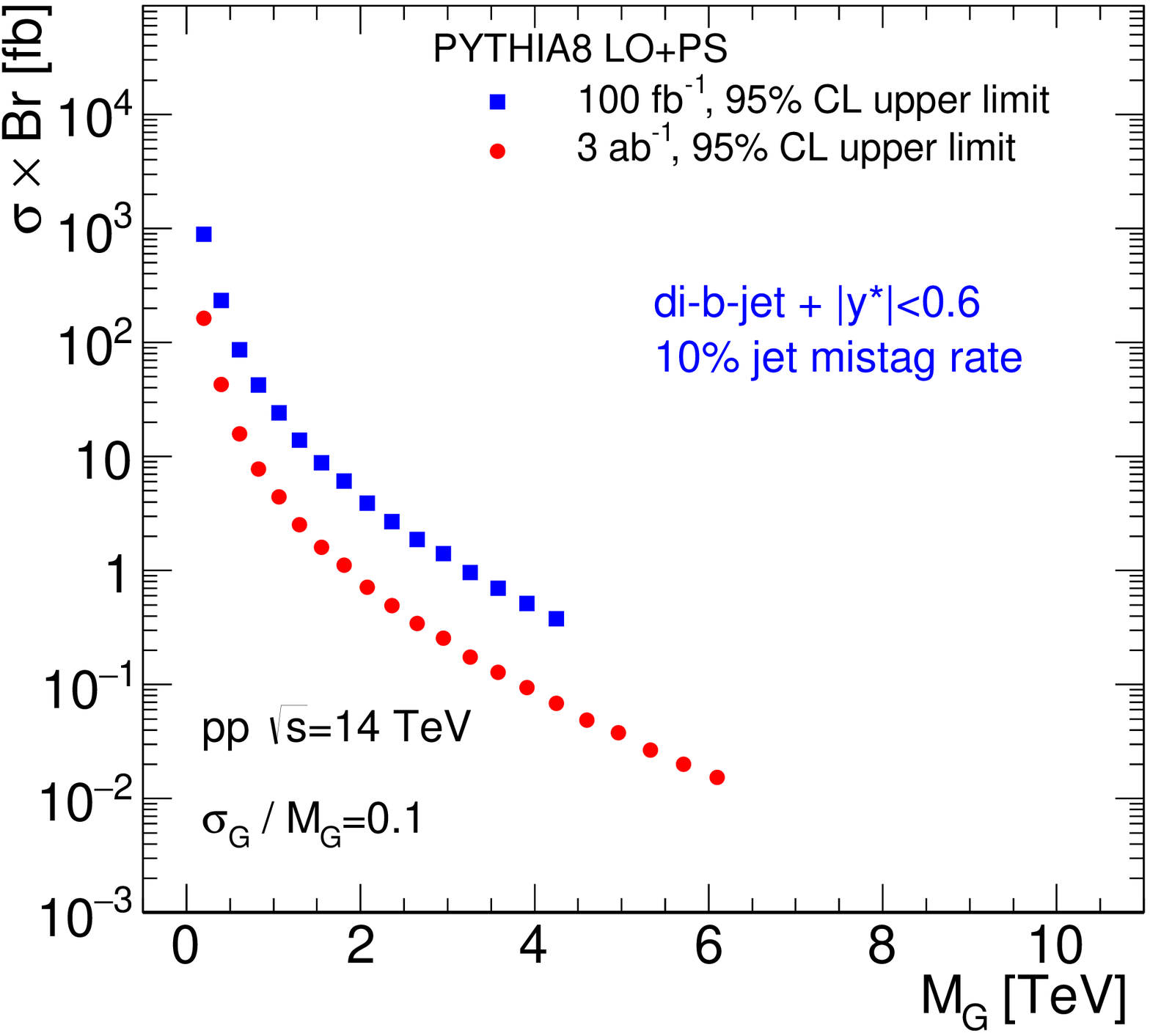}\hfill
   }
\end{center}
\caption{The 95\% C.L. upper limits obtained from the
$\mjj$  distribution on cross section times the  branching ratio to two jets (identified as $b$-jets) 
for a hypothetical signal approximated by a Gaussian whose width is 10\% of the mass of the searched resonance. 
The limits are shown (a) without and (b) with $y^*$ cut.}
\label{fig:14tev_limits_JetJetMass_2bjet}
\end{figure}

Figure~\ref{fig:14tev_limits_JetJetMass_2bjet} shows the $95\%$ C.L. upper
limit for the cross section times the branching ratio 
for a signal 
approximated by a Gaussian whose width is 10\% of the mass of the searched resonance. 
In addition to the dijet masses, we also calculate the upper limits 
after applying the rapidity difference requirement $|y^*|<0.6$ between the 
two jets \cite{ATLAS:2015nsi,2016229}. 

In order to calculate the expected upper limits for observation of   
particles such as $Z'$,
we have performed a simulation of $Z'$ decays to $b$-jets, assuming that its width is $15\%$.  
Exclusion limits were also calculated using the $CL_{s}$ method with a binned profile likelihood ratio as the test statistic using the {\sc HistFitter}   framework.
Figure~\ref{fig:14tev_limits_JetJetMass_2bjet} shows the  95\% C.L. upper for the realistic signal shapes. 

Several BSM models, such as $b^*$  and leptophobic $Z'$, predict peaks in the $\mjj$ distribution, where  
one or two jets are identified as $b$-jet.  Some models have already been excluded by ATLAS~\cite{2016229}.

\begin{figure}[h]
\begin{center}
   \subfigure[ $100~\fbb$] {
   \includegraphics[width=0.45\textwidth]{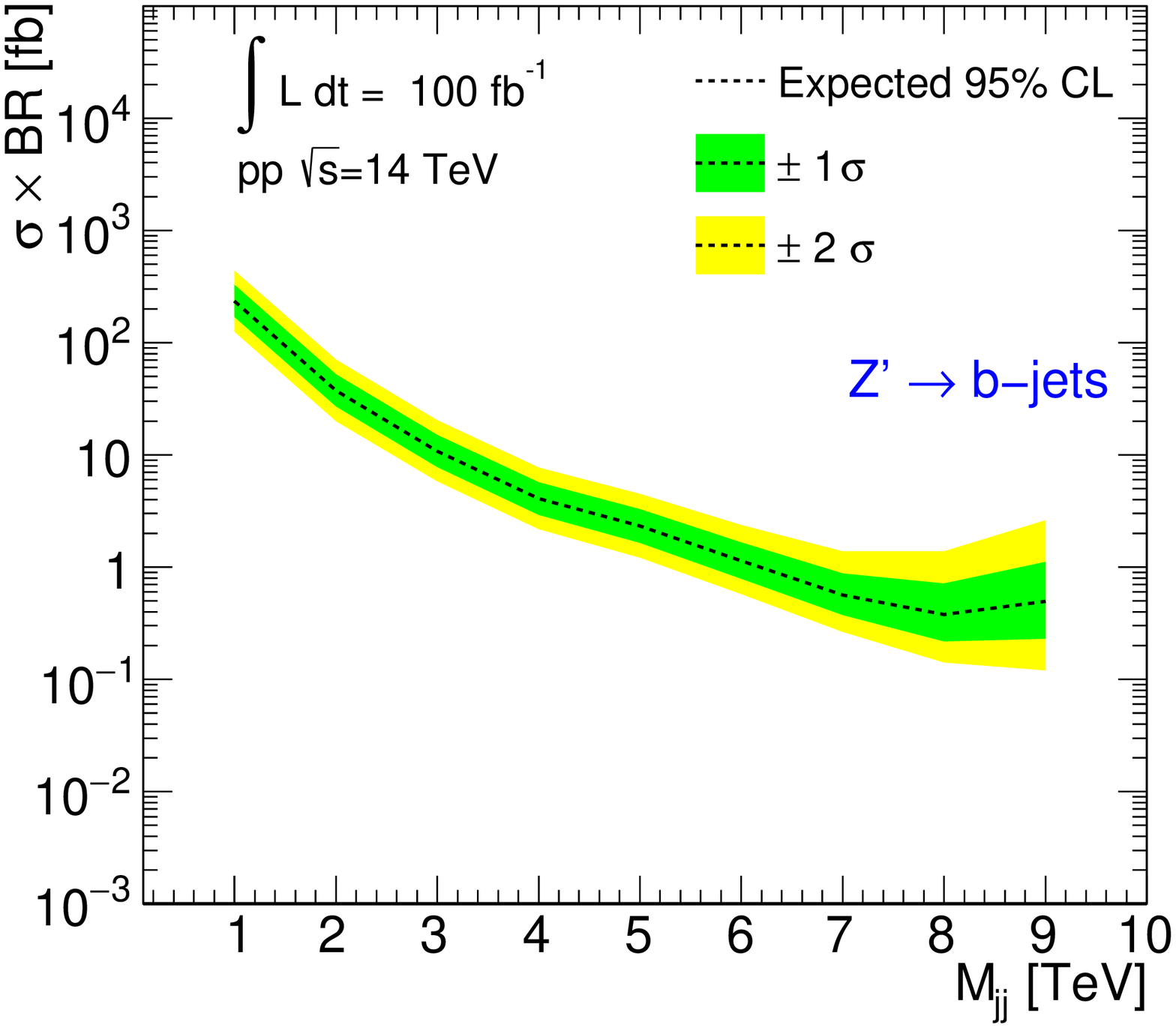}\hfill
   }
   \subfigure[ $3~\abb$] {
   \includegraphics[width=0.45\textwidth]{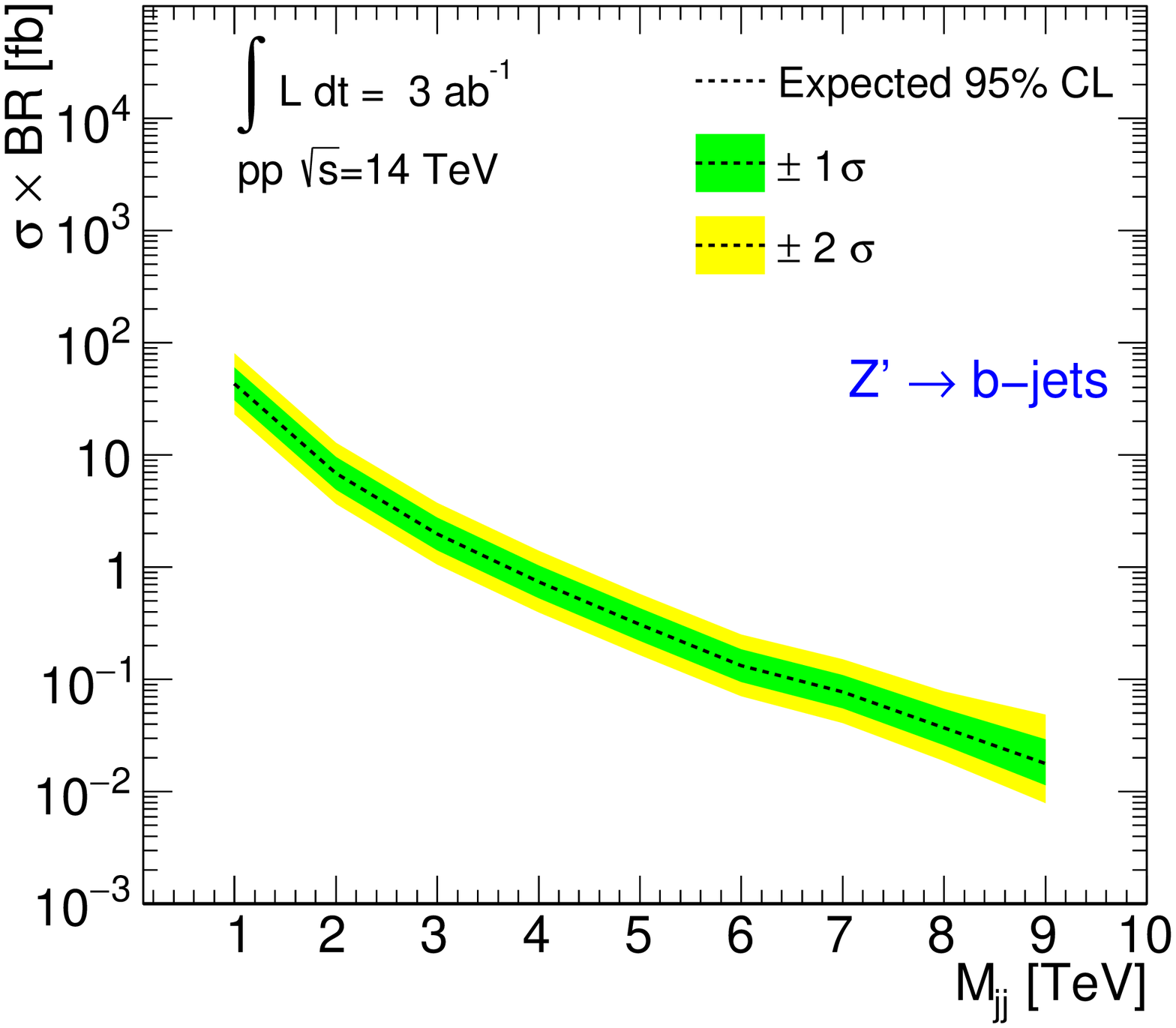}
   }
\end{center}
\caption{The 95\% C.L. upper limits obtained from the $\mjj$  distribution on cross section times the  branching ratio to two jets for $Z'$ particle decaying to two $b$-jets. }
\label{fig:zprime_limits_JetJetMass_2bjet}
\end{figure}
\clearpage

\begin{figure}[h]
\begin{center}
   \subfigure[ $100~\fbb$] {
   \includegraphics[width=0.45\textwidth]{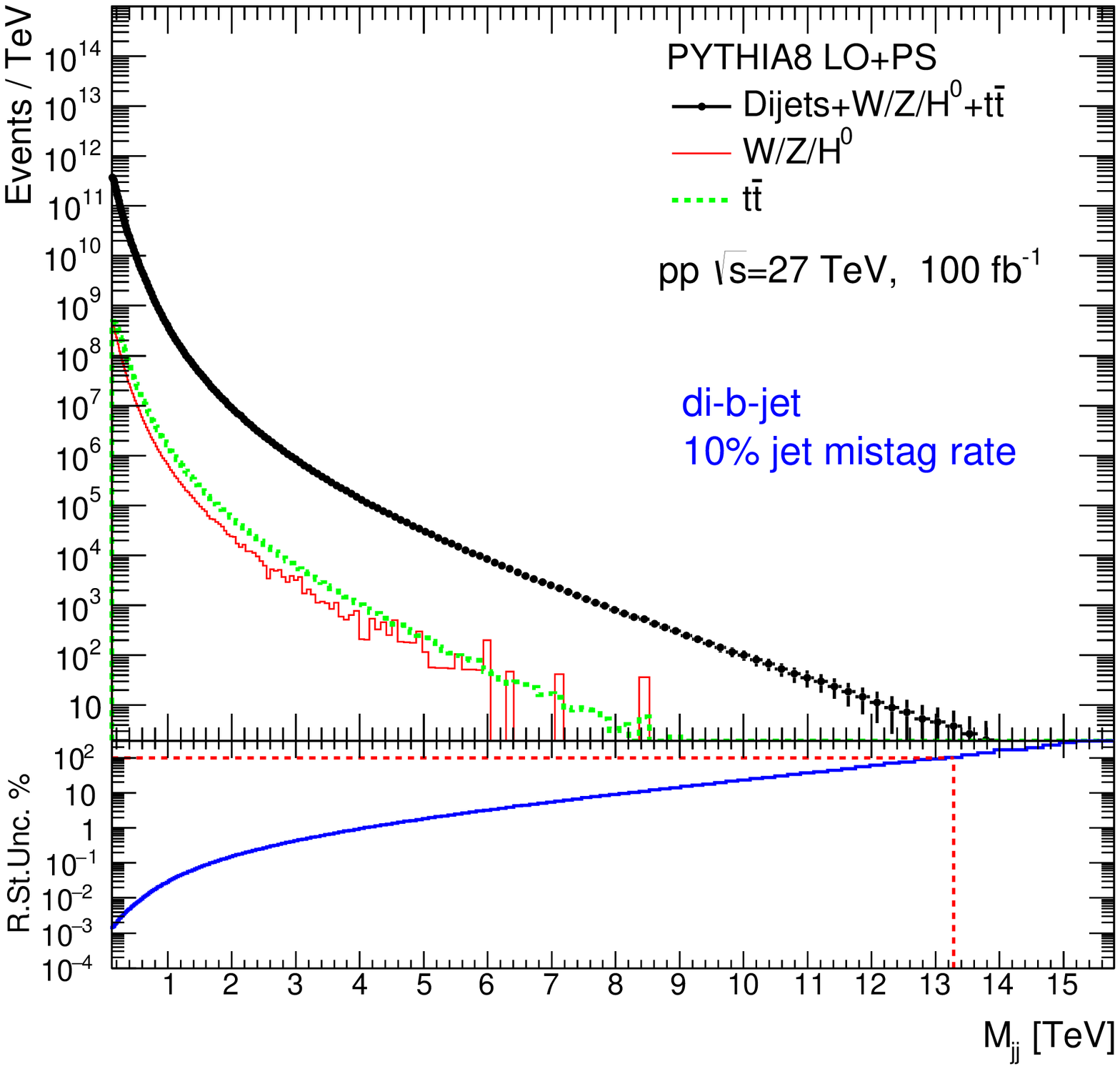}\hfill
   }
   \subfigure[ $3~\abb$] {
   \includegraphics[width=0.45\textwidth]{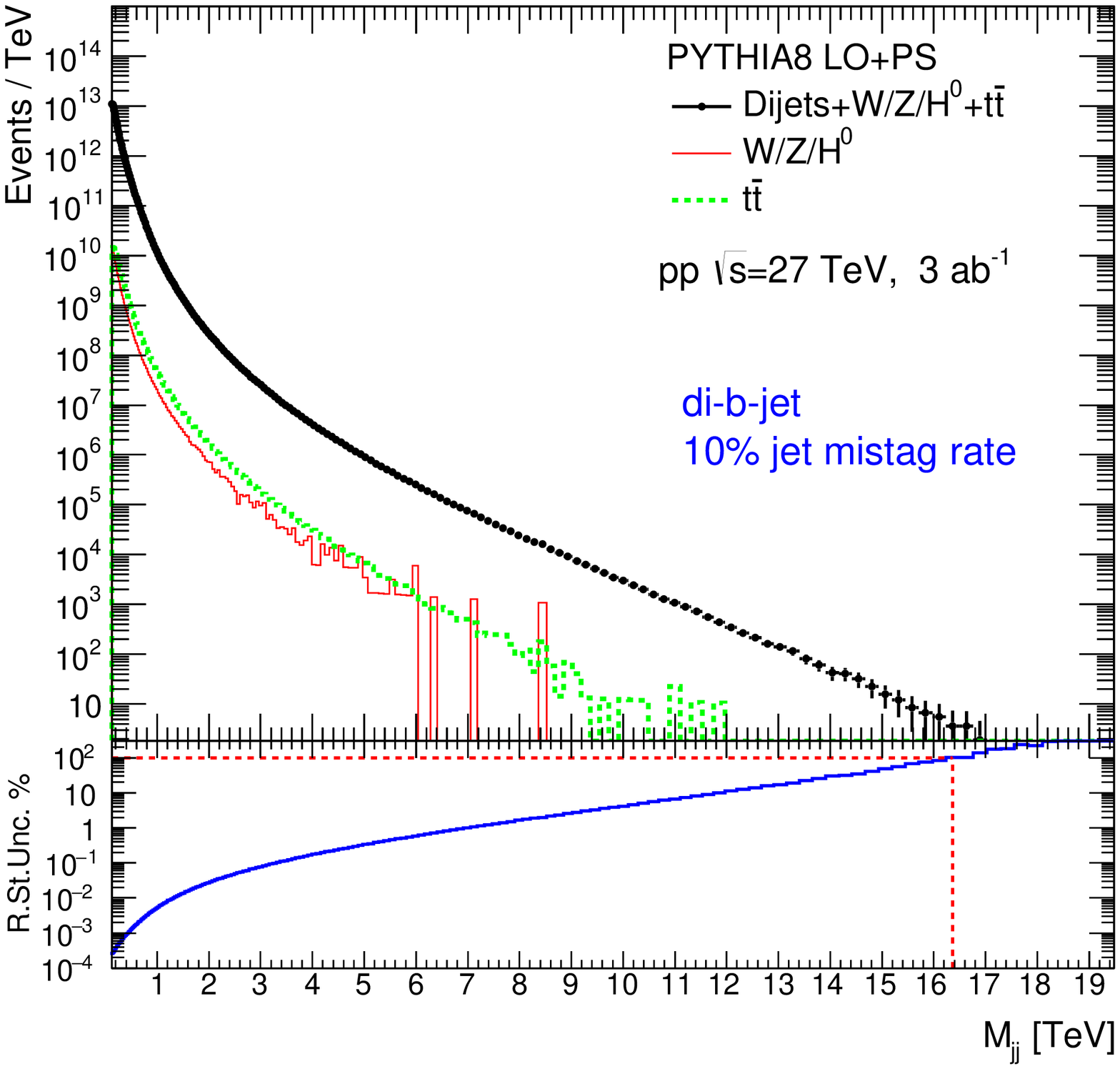}
   }
\end{center}
\caption{
Dijet mass distributions for jets identified as $b$-jets,
for 100~$\fbb$ and 3~$\abb$ at the HE-LHC.}
\label{fig:27tev_JetJetMass_2bjet}
\end{figure}

\begin{figure}[h]
\begin{center}
   \includegraphics[width=0.5\textwidth]{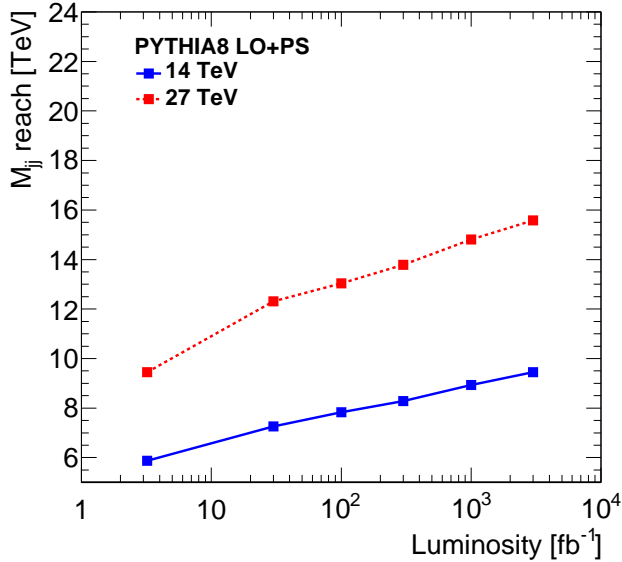}\hfill
\end{center}
\caption{Dijet mass reach for the HL-LHC and HE-LHC experiments for dijets identified as $b$-jets. The uncertainties on the data
points, derived from the bin width used for the simulated $\mjj$ distributions, are compatible with the size of the symbols.}
\label{fig:mass_reach_dib}
\end{figure}

The \pythia expectations for the $\mjj$ distribution with jets identified as 
$b$ jets at the  HE-LHC collider are shown in Fig.~\ref{fig:27tev_JetJetMass_2bjet}.
The line on the lower panel indicates  the mass at which the relative statistical uncertainty is  $100\%$ on the data point.
As before, this point is chosen to define the dijet  mass reach to be accessible for the given luminosity.
This point may depend on the assumption used for the fake rate and the efficiency discussed  in Sect.~\ref{sec:reco},
but the difference between different luminosity scenarios should not depend much on these assumptions. 
Figure~\ref{fig:mass_reach_dib} shows  the mass reach for the HL-LHC and HE-LHC. 
For the  modest luminosity of $100~\fbb$, the mass reach at the HE-LHC is above 13~TeV,
which is a factor of two larger than for the HL-LHC, assuming that the reconstructed efficiencies and fake rates 
for $b$-jets are similar for the HE-LHC and HL-LHC.

\begin{figure}[h] 
\begin{center}
   \subfigure[ No $y^*$ cut] {
   \includegraphics[width=0.45\textwidth]{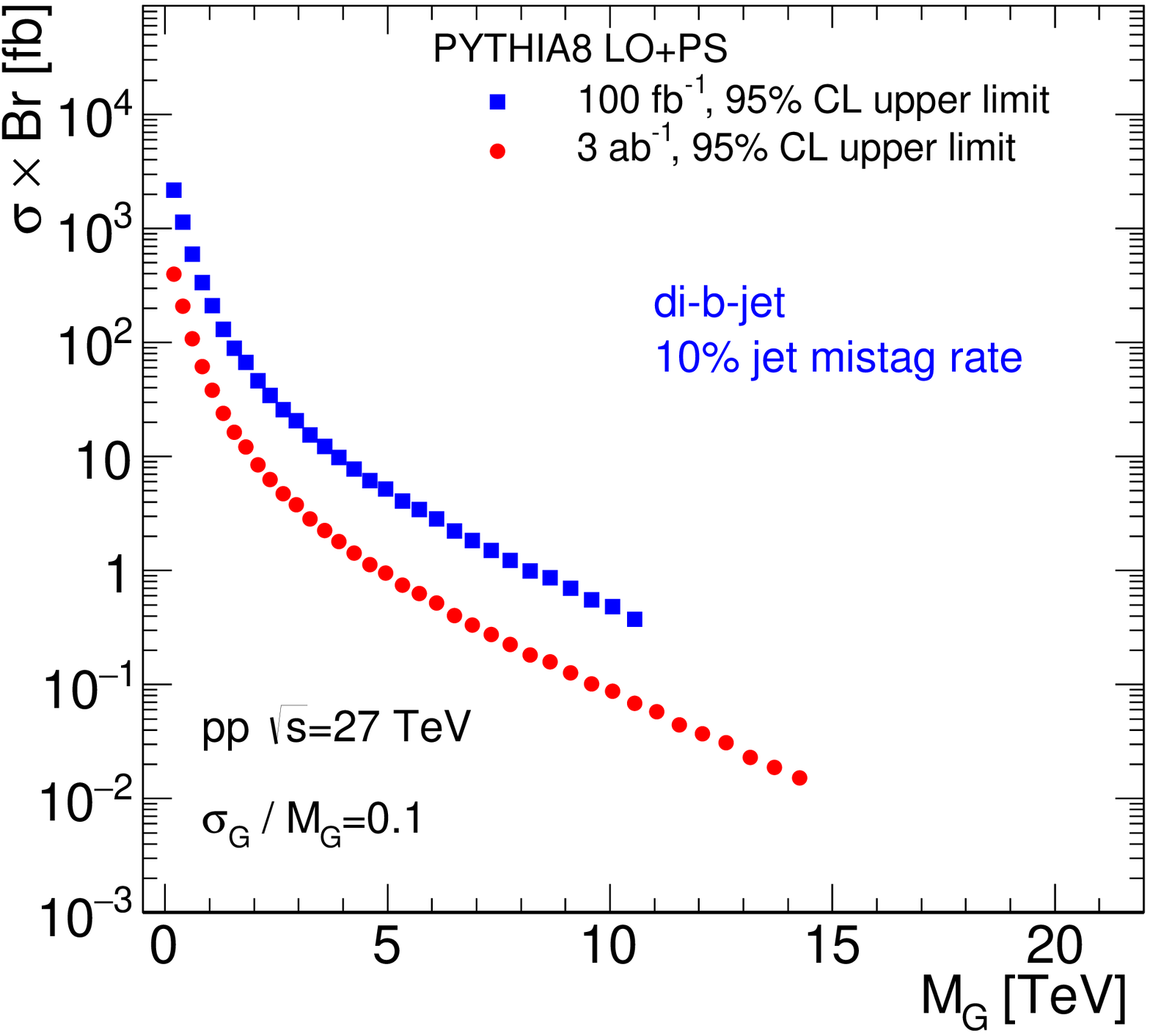}\hfill
   }
   \subfigure[ With $y^*$ cut] {
   \includegraphics[width=0.45\textwidth]{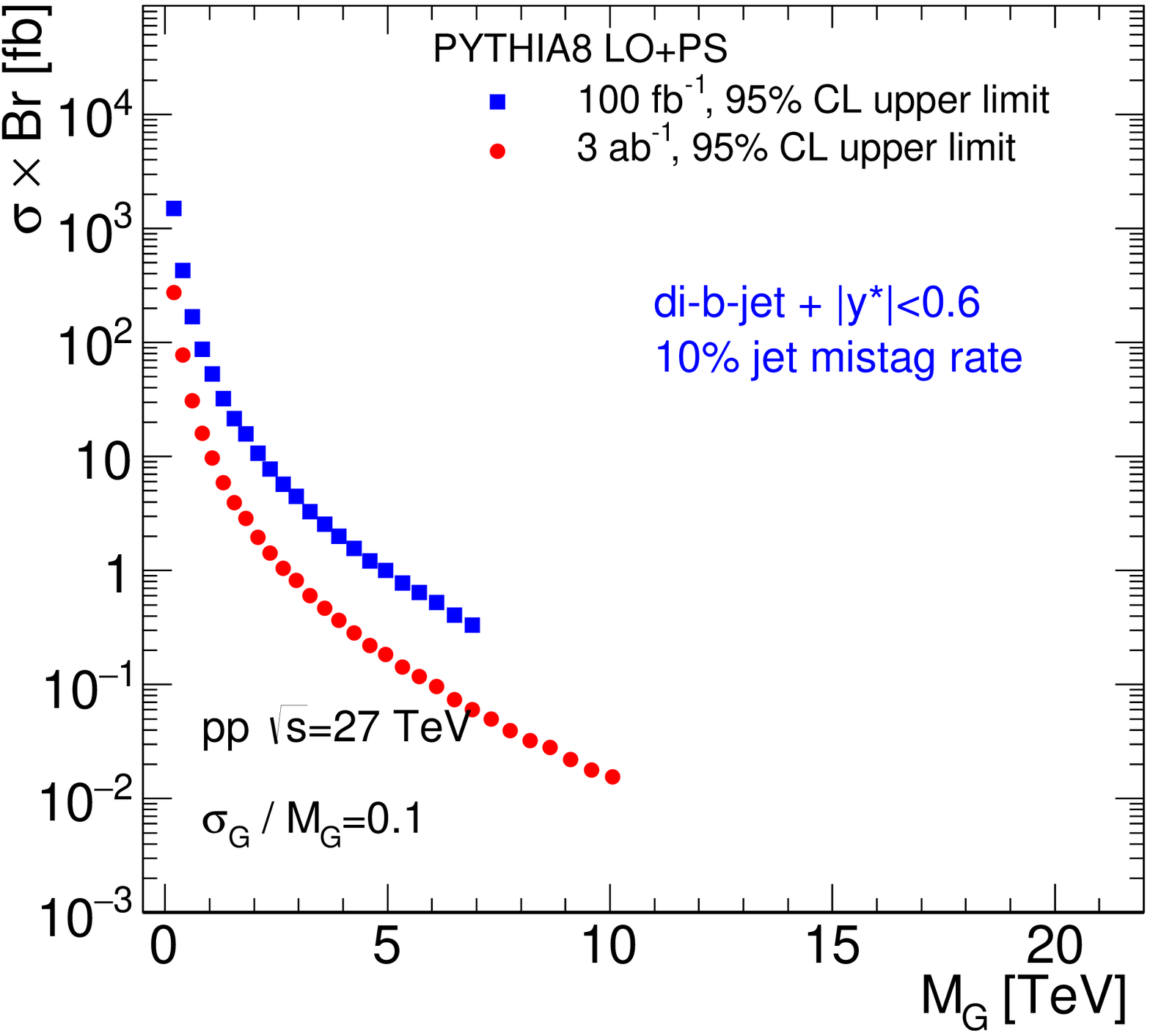}\hfill
   }
\end{center}
\caption{The 95\% C.L. upper limits obtained from the $\mjj$  distribution on cross section times the  branching ratio to two jets (identified as $b$-jets) for a hypothetical signal approximated by a 
Gaussian contribution to the expected dijet mass. The results are shown for the HE-LHC (without and with $y^*$ cut).}
\label{fig:27tev_limits_JetJetMass_2bjet}
\end{figure}

Figures~\ref{fig:27tev_limits_JetJetMass_2bjet} shows 95\% C.L.  upper limits for the cross section times the branching ratio
for a signal approximated by a Gaussian with $\sigma_G=0.1\cdot M_{G}$.

\begin{figure}[h]
\begin{center}
   \subfigure[ $100~\fbb$] {
   \includegraphics[width=0.34\textwidth]{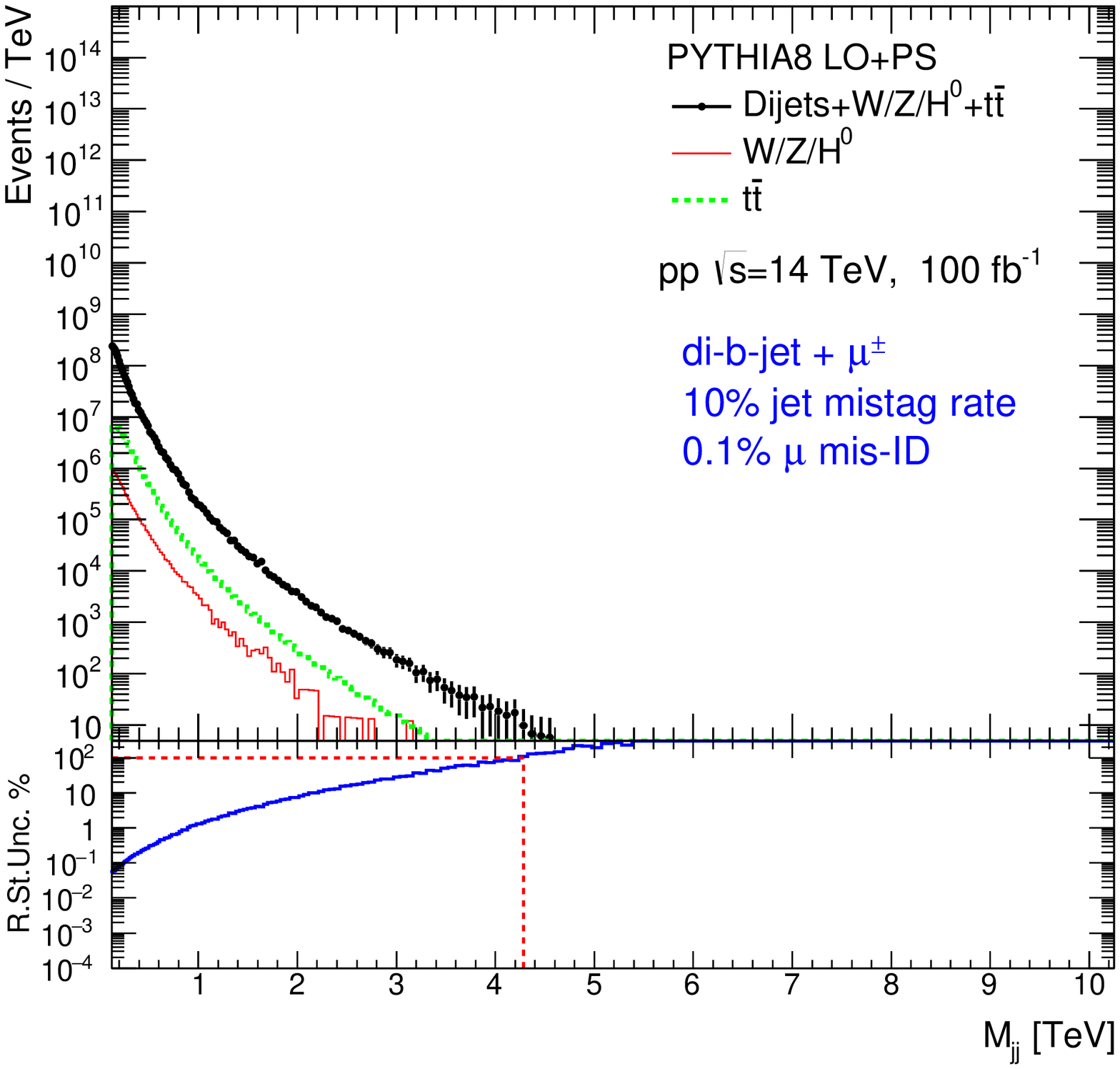}\hfill
   }
   \subfigure[ $3~\abb$] {
   \includegraphics[width=0.34\textwidth]{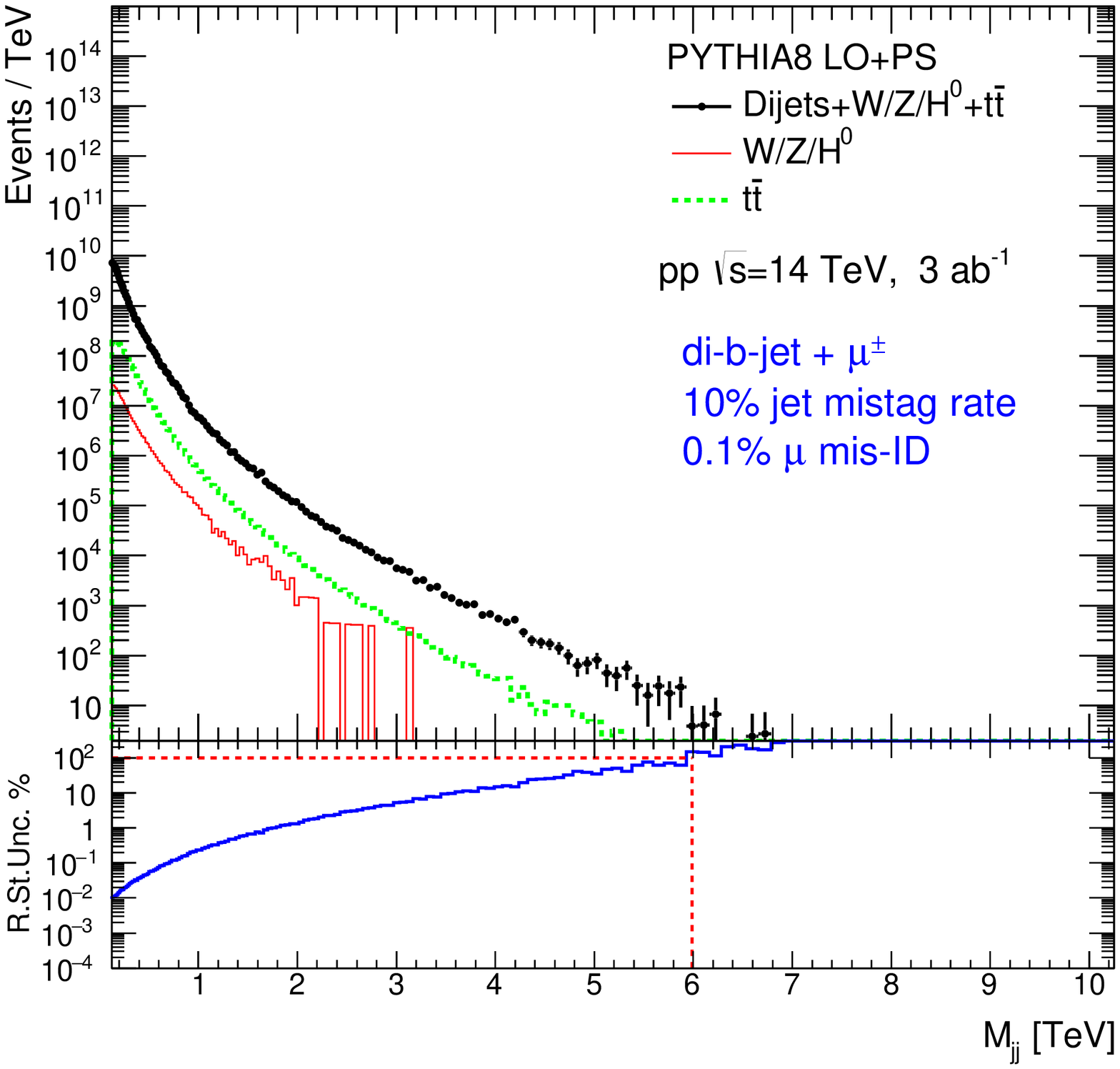}
   }
   \subfigure[ $100~\fbb$] {
   \includegraphics[width=0.34\textwidth]{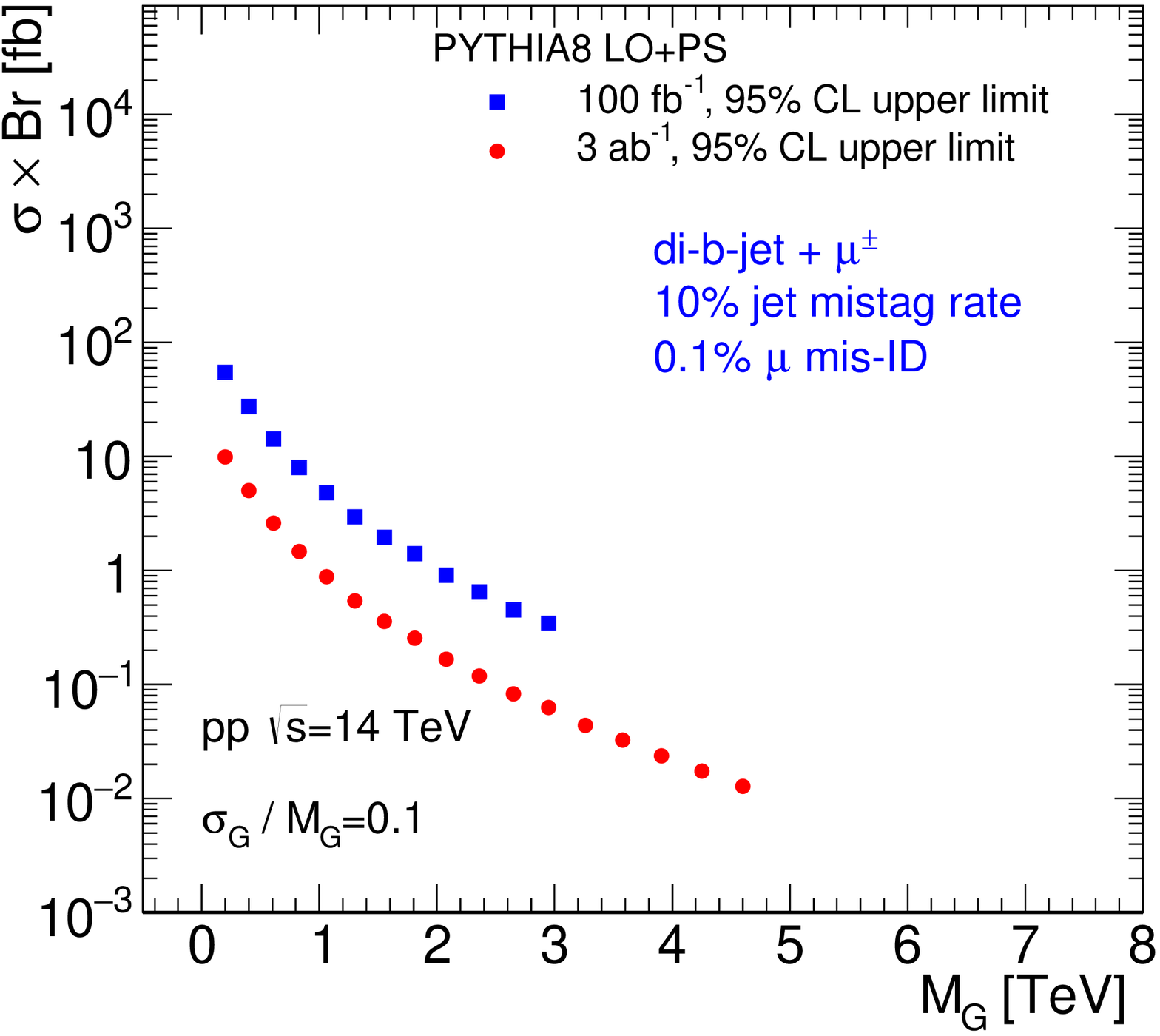}\hfill
   }
   \subfigure[ $3~\abb$] {
   \includegraphics[width=0.34\textwidth]{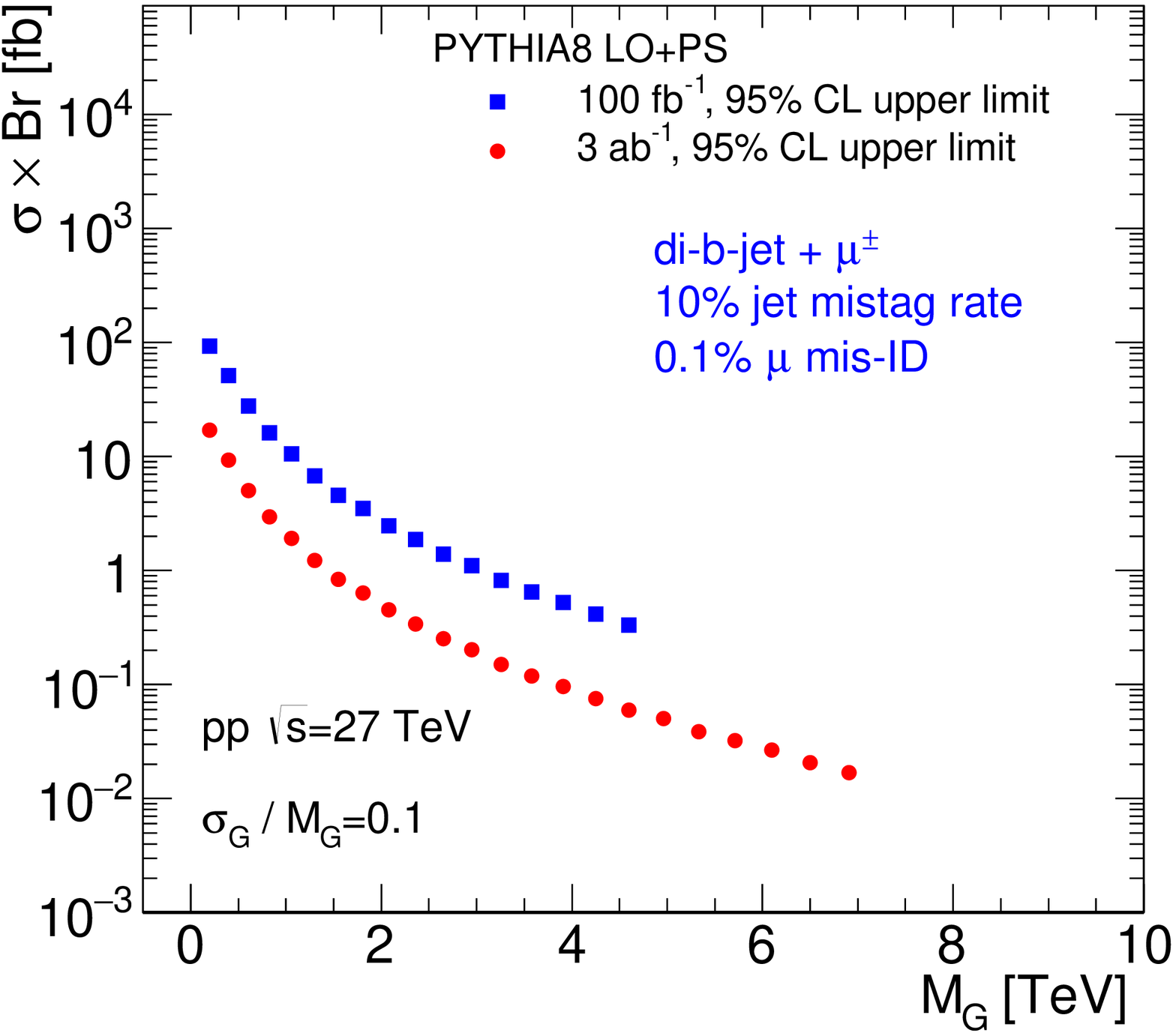}
   }
  \subfigure[ $100~\fbb$] {
   \includegraphics[width=0.34\textwidth]{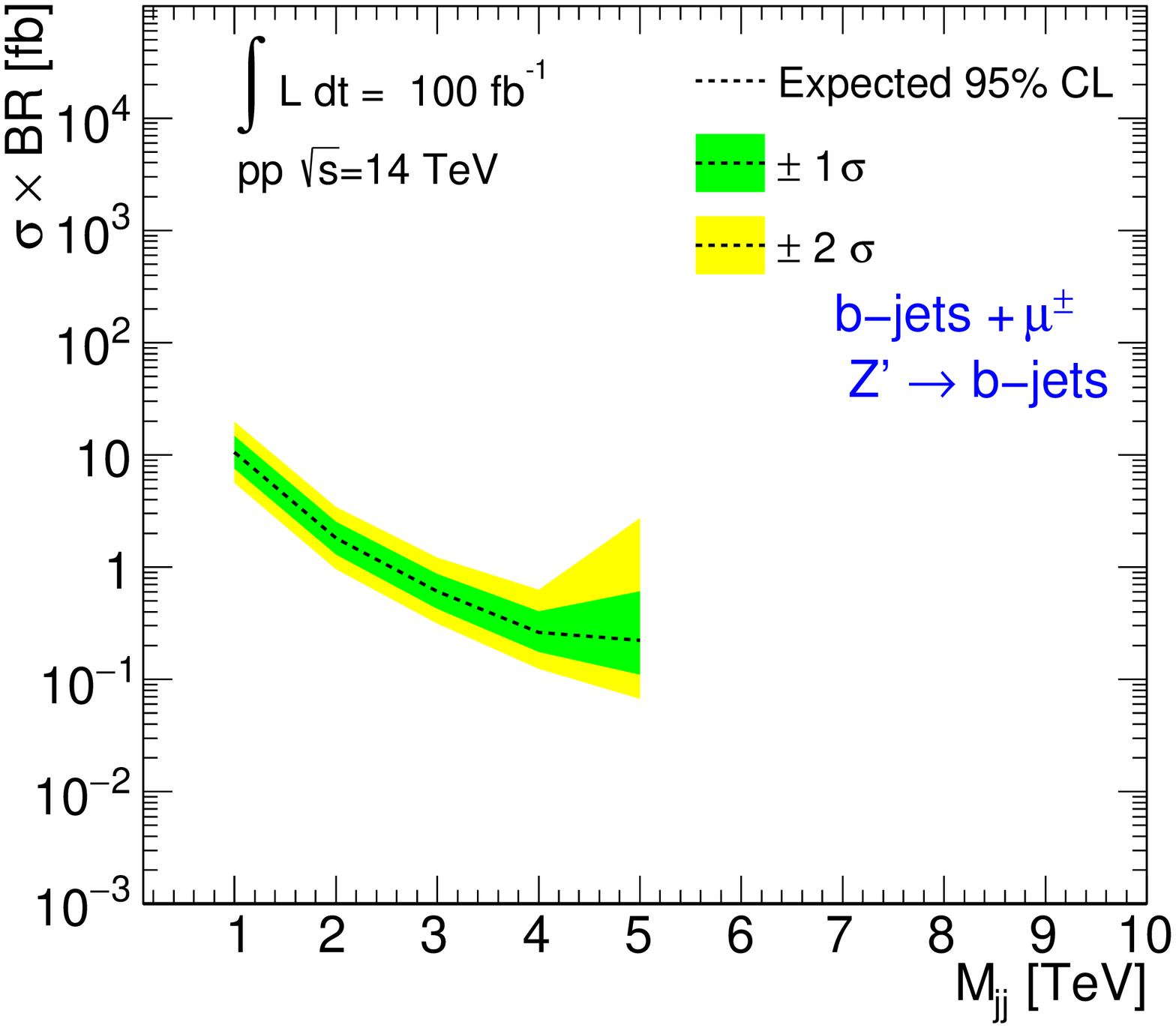}\hfill
   }
   \subfigure[ $3~\abb$] {
   \includegraphics[width=0.34\textwidth]{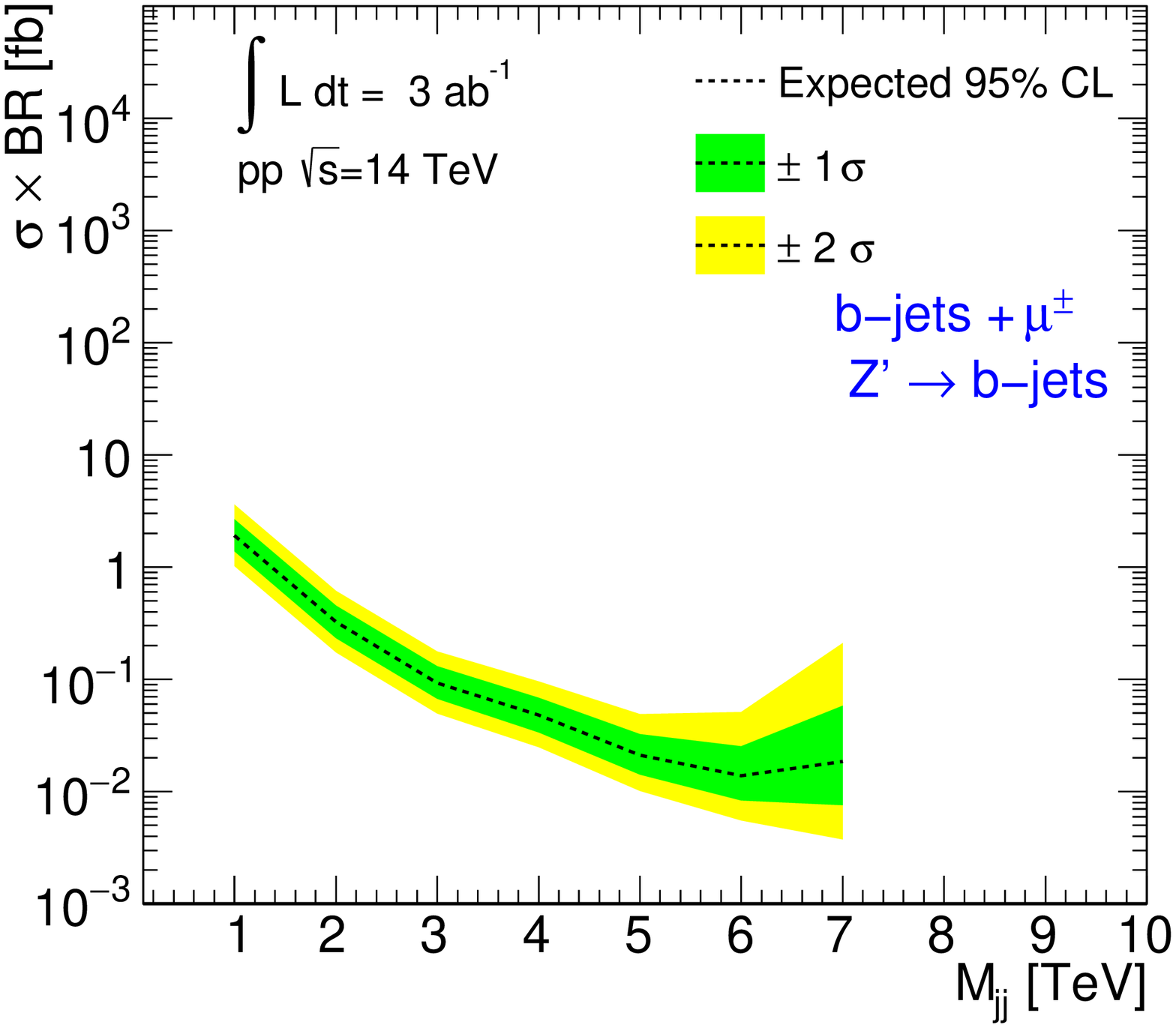}
   }

\end{center}
\caption{(a)-(b) show the  distribution of dijet invariant masses  with jets  identified as $b$-jets,
in events with associated muons for 100~$\fbb$ and 3~$\abb$.
(c)-(d) 95\% C.L. upper limits obtained from the
on cross section times the  branching ratio to two jets
for a hypothetical signal approximated by a Gaussian contribution to the expected dijet mass. The limits
are obtained for events with isolated muons.
(e)-(f) 95\% C.L. upper limits for a $Z'$ particle decaying to two jets for the centre-of-mass energy of 14~TeV.}
\label{fig:14tev_JetJetMass_2bjet_mu}
\end{figure}

Now we will consider events with $b$-jets measured in events with   
an  additional identified lepton.
Figure~\ref{fig:14tev_JetJetMass_2bjet_mu}(a)(b) shows the dijet invariant masses with isolated muons having 
$p_T>60$~GeV. 
The total event rate is reduced to 2\% of the $b$-tagged dijet. It can also be noted that the contribution from $W/Z/H^0$ is at the level of $1.2\%$ while the $t$-quark processes are 
at the level of $3\%$, which is a factor of ten larger than the contribution shown in 
Fig.~\ref{fig:14tev_JetJetMass_2bjet} of Sect.~\ref{sect_bjets}.  

For a completeness, Fig.~\ref{fig:14tev_JetJetMass_2bjet_mu}(c)(d) show 
the 95\% C.L.  upper limits for the cross section times the branching ratio 
for a generic  Gaussian signal, assuming    
the muon-associated dijet production.
Figure~\ref{fig:14tev_JetJetMass_2bjet_mu}(e)-(f) show the  95\% C.L. upper limits for 
a $Z'\rightarrow b\bar{b}$ signal.
Generally, the exclusion limits  at a fixed mass obtained  using 3~$\abb$ are improved by a factor 10 compared to the 100~$\fbb$ case.

%%%%%%%%%%%%%%%%%%%%%%%%%%%%%%%%%%%%%%%%%%%%%%%%%%%%%%%%%%%%%%%%%%
\section{Signal extraction}
\label{sect_signal}
%%%%%%%%%%%%%%%%%%%%%%%%%%%%%%%%%%%%%%%%%%%%%%%%%%%%%%%%%%%%%%%%%%

The use of an associated lepton in the event selection allows a uniform exploration of the dijet mass distribution 
which spans 14 orders of magnitude in rate. 
The natural question arises how to extract features that may correspond to 
signal events from BSM physics. 
To illustrate the difficulties arising in a data-driven signal extraction,  we will consider an analytic fit 
of  dijet mass spectra with the monotonically decreasing function: 
\begin{equation}
  f(x) = p_1 (1 - x)^{p_2} x^{p_3 + p_4\ln x},
\label{eq:function}
\end{equation}
were $x = \mjj /\sqrt{s}$ and $p_i$\ are 
fit parameters. This function was used for inclusive dijet searches \cite{2016229,Aaboud:2017yvp,Sirunyan:2016iap}
by both ATLAS and CMS collaborations. 

\begin{figure}[h]
\begin{center}
   \subfigure[Dijets at 14 TeV] {
   \includegraphics[width=0.45\textwidth]{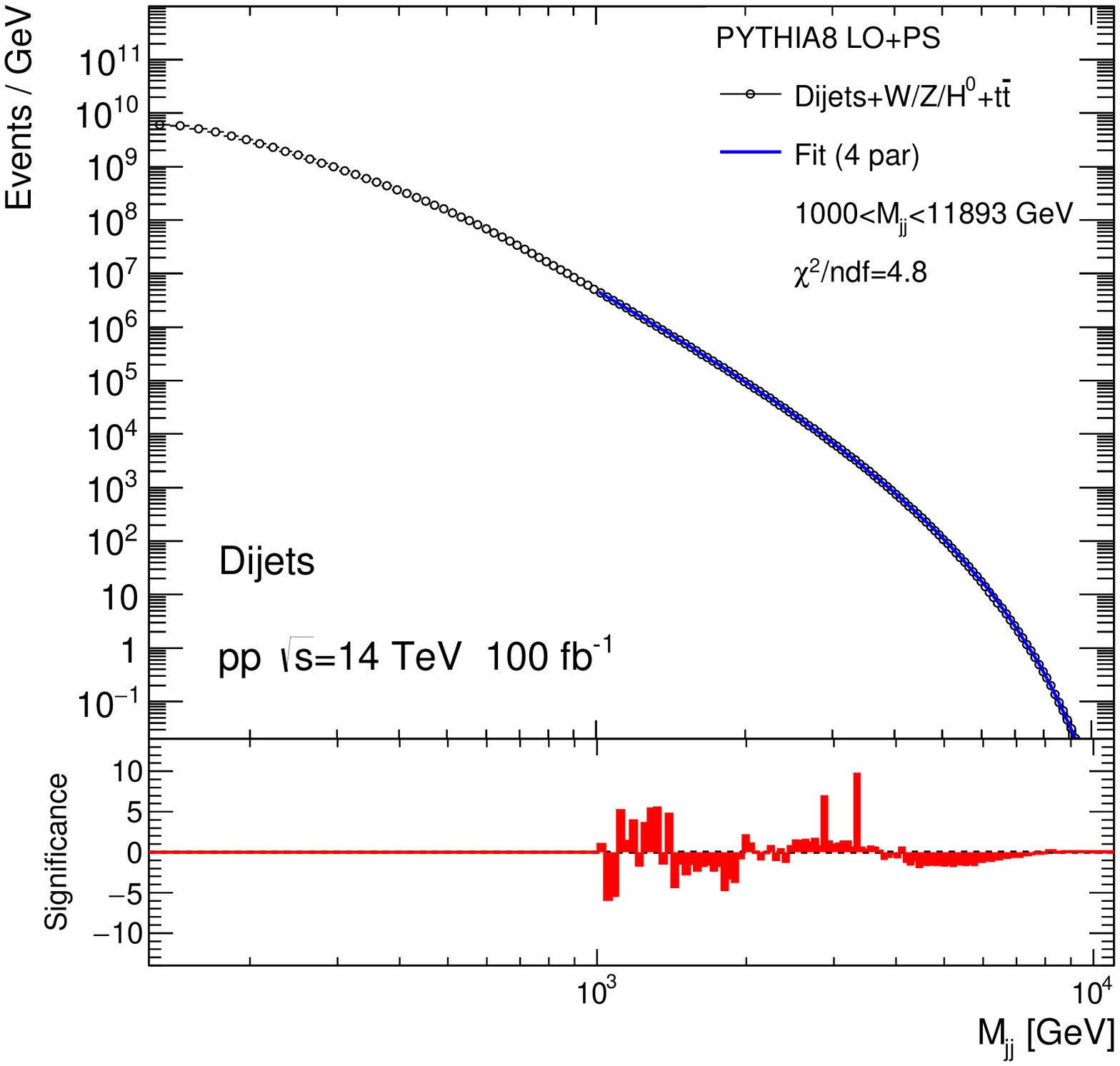}\hfill
   }
   \subfigure[Dijets plus muon at 14 TeV] {
   \includegraphics[width=0.45\textwidth]{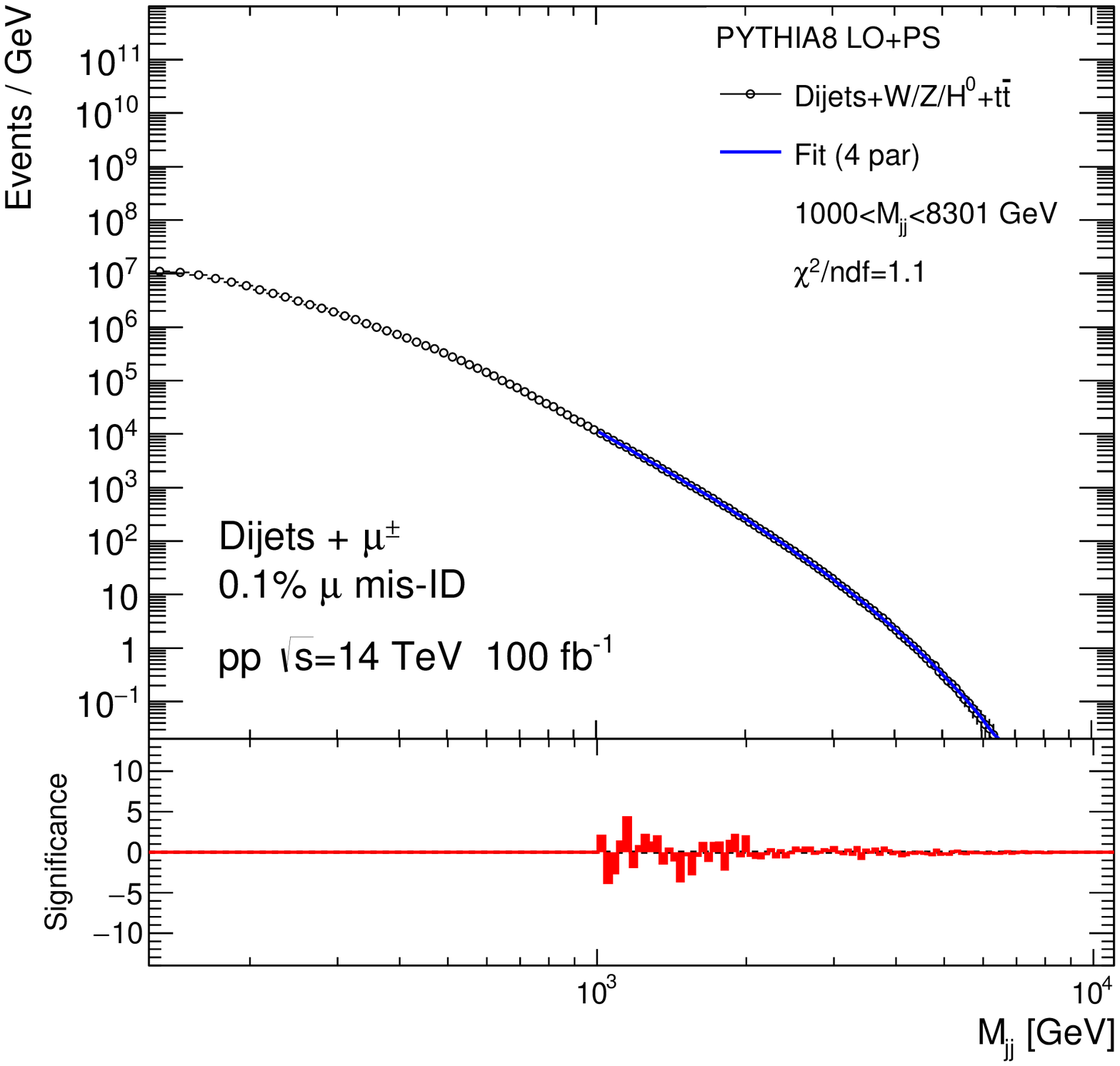}
   }
\end{center}
\caption{Dijet invariant masses shown together with
the analytic fit function Eq.~\ref{eq:function} after the $\chi^2$ minimization.
The simulations were performed for (a) inclusive dijet events and (b) for dijet events with associated muon. 
The results are obtained using  $100~\fbb$.}
\label{fig:signal_JetJetMass_2jet_100fb}
\end{figure}

Figure~\ref{fig:signal_JetJetMass_2jet_100fb} shows the mass of two jets  
together with the fit of Eq.~\ref{eq:function}. The distribution 
for jets associated with isolated muons is shown in Fig.~\ref{fig:signal_JetJetMass_2jet_100fb}(b).
The bottom plot shows the significance of deviations of the function from simulated data in terms
of the variable $S_i= (d_i-f_i) / \Delta d_i$, where $d_i$ is the simulated data point in a bin $i$,
$f_i$ is the value of the function after the $\chi^2$ minimization, and
$\Delta d_i$ is the statistical uncertainty on the value of $d_i$.
We consider the fit scenario when the function Eq.~\ref{eq:function} is applied to the dijet mass spectrum
below $1$~TeV, and above $1$~TeV, separately.

\begin{figure}[h]
\begin{center}
   \subfigure[14 TeV, $3~\abb$, $125<M_{jj}^{fit}<1$ TeV] {
   \includegraphics[width=0.45\textwidth]{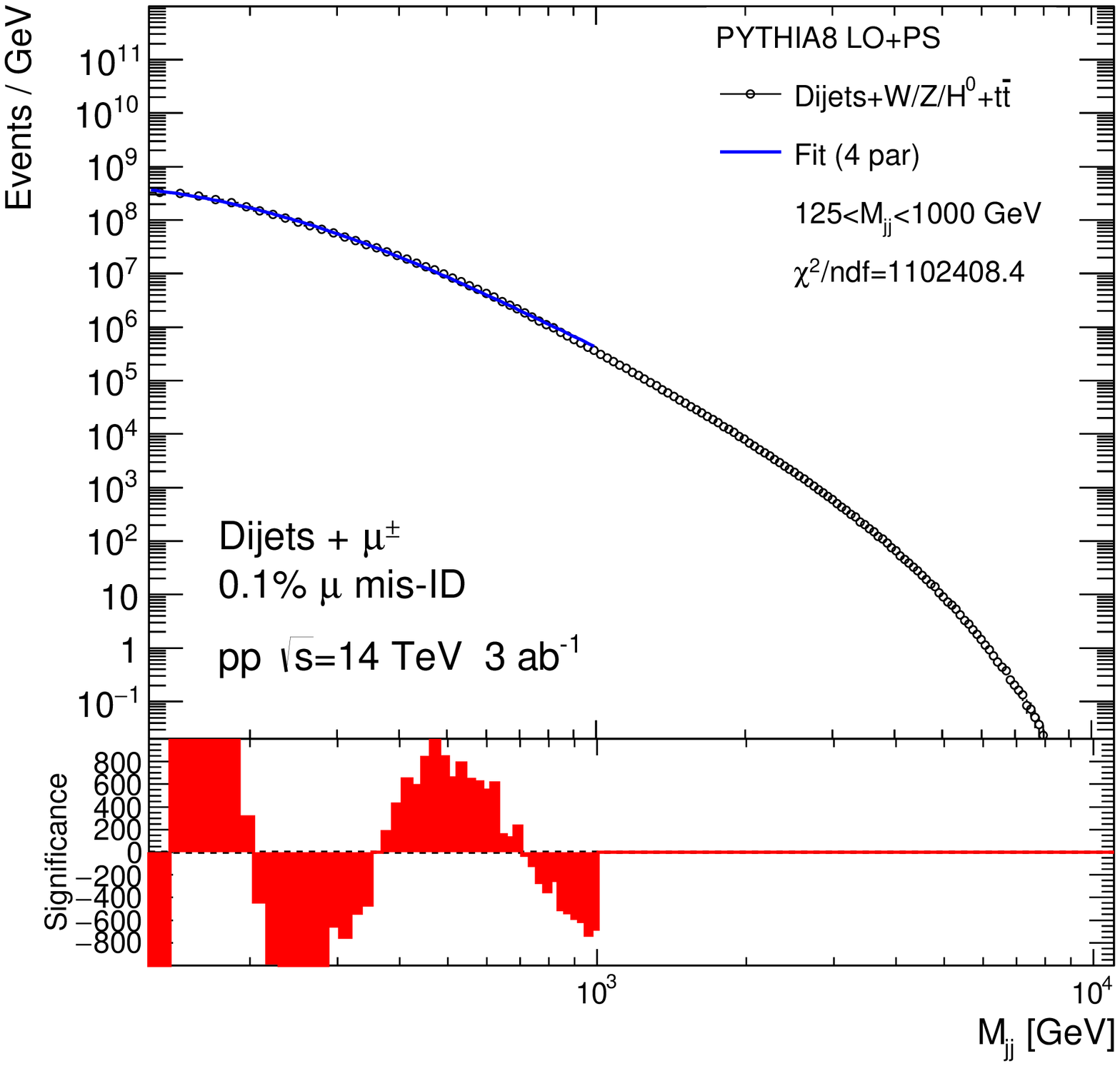}\hfill
   }
   \subfigure[14 TeV,  $3~\abb$, $M_{jj}^{fit}>1$ TeV] {
   \includegraphics[width=0.45\textwidth]{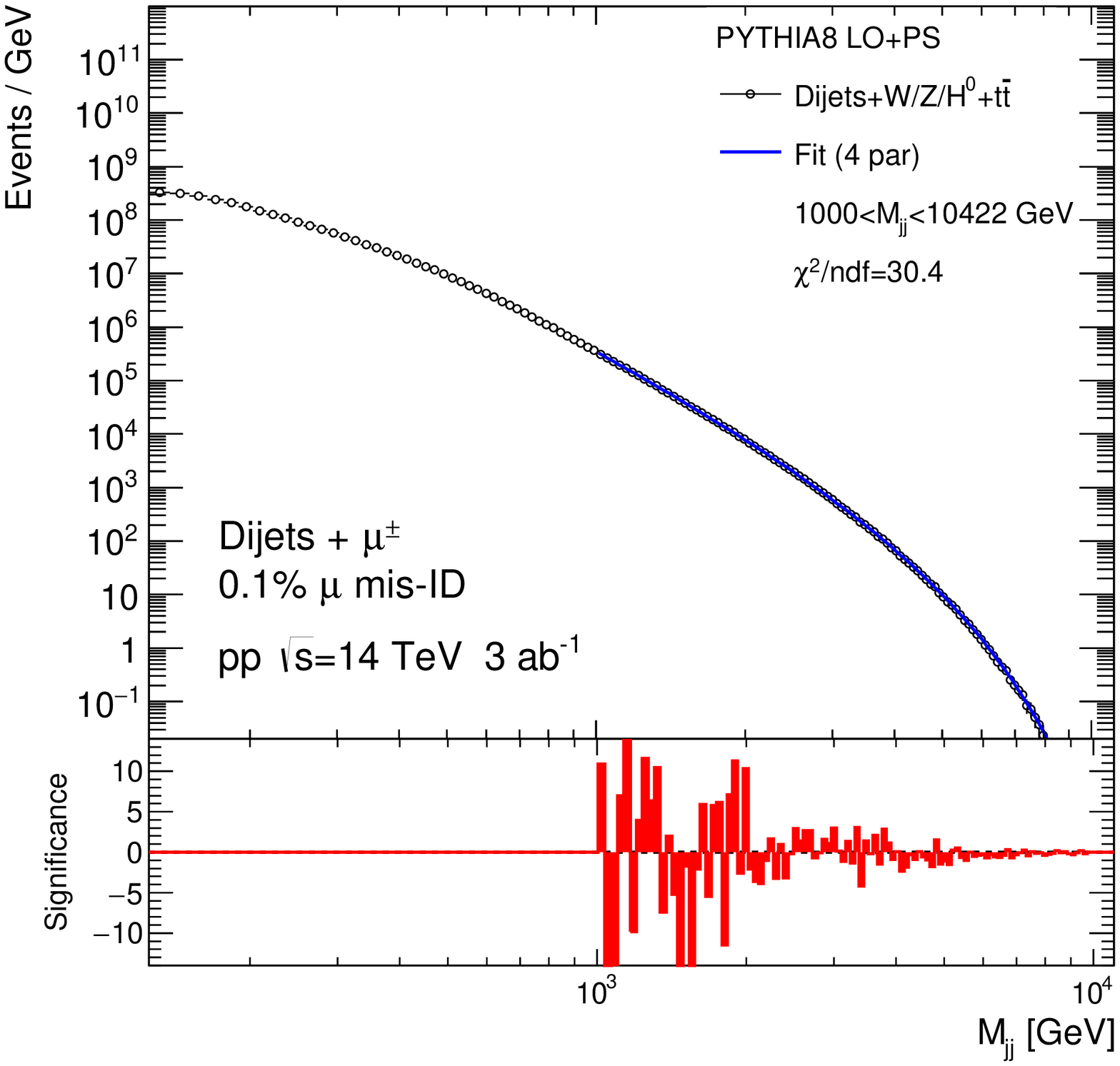}
   }
\end{center}
\caption{Dijet invariant masses in events with muons with $p_T>60$~GeV fitted with the
analytic function Eq.~\ref{eq:function} for different mass ranges.
The simulations are shown for the HL-LHC collider.}
\label{fig:signal_JetJetMass_2jet_mu}
\end{figure}

The fit function reasonably describes the mass distribution above $1$~TeV. A similar good agreement 
between data and the fit function was shown  
in the previous studies \cite{2016229,Aaboud:2017yvp,Sirunyan:2016iap}  that used smaller integrated luminosities.
When using the same integrated luminosity as in Ref.~\cite{2016229,Aaboud:2017yvp,Sirunyan:2016iap},
no statistical deviations from the fit function were found. The quality of the fit was given by $\chi^2/ndf=0.9$.

Figure~\ref{fig:signal_JetJetMass_2jet_mu} shows the mass of two jets associated with isolated muons together
with the fit of Eq.~\ref{eq:function} for the HL-LHC case.  In addition to the high-mass region, the figure
shows the fit results for $0.125<\mjj<1$~TeV.  
This scenario is relevant for semi-inclusive searches using an additional 
trigger requirement 
on leptons, and other triggered particles.   
The function does describe the low-mass and high-mass regions in \pythia, 
even when the fit was performed separately in these two mass windows.
This can be seen from the quoted $\chi^2/ndf$ value, and by observing oscillations of $S_i$. 

As a test, we have performed a $\chi^2$ minimization 
with Eq.~\ref{eq:function} using the $\mjj$ distribution for inclusive dijets in different $\mjj$ regions.
To make sure that the observed feature does not come from the phase-space re-weighting (see Sect.~\ref{sect_mc}),
the simulation was performed without the re-weighting, but using a smaller sample of events. 
Similar oscillating values of $S_i$ were found.
It was found that the fit cannot describe the entire mass spectra, $0.125<\mjj<1$~TeV, where 
it significantly underestimates the tail
of the distribution.
The fit also fails for the HE-LHC energies (not shown).
A similar behavior in the fit was observed after multiplying  the fit function by the additional term $x^{p_5\ln^2 x}$.

Additional studies have been conducted by taking numerical derivatives of the 
function Eq.~\ref{eq:function}. 
It was found that the first derivatives of the function and data are monotonically increasing, while the 
second derivatives are always positive and do not cross the value zero. 
Therefore, the oscillatory behavior of $S_i$ after the $\chi^2$ minimization is a reflection
of differences between the  shape of the $\mjj$ distribution in \pythia and the analytic function,
rather than a consequence of the presence of oscillations in the simulated data.

Our results show that there are many challenges at the HL-LHC and HE-LHC 
in data driven methods to extract signals
in the bulk of $\mjj$ distribution, where the relative statistical uncertainties 
in $\mjj$ bins reach $0.01\%$. The analytic approach based on Eq.~\ref{eq:function} cannot describe the $\mjj$  
observed in \pythia.  Therefore, small features  in the form of peaks from BSM  physics 
can be masked by the oscillatory behavior of the fit residuals shown in Fig.~\ref{fig:signal_JetJetMass_2jet_mu}.
In addition to the fit function technique, numerical  techniques for data-driven signal extraction may also be used,  assuming  
they can reliably describe the shape for multijet QCD background and, at the same time, are sensitive
to the presence of small peaks in dijet mass distributions. 
In the past, a  number of peak-identification and data smoothing algorithms were  proposed for counting-type observables 
\cite{Huang1969141, *Mariscotti1967309, *Blinowska1974597, *Silagadze:1995zd, *2000NIMPA.443..108M, *Chekanov:2011vj,*Frate:2017mai}.
An approach based on  a sliding-window fitting
technique was used by ATLAS \cite{Aaboud:2017yvp} to overcome the complexity of $\mjj$ distributions.  
%It would be interesting to verify such techniques using the simulated data presented in this paper.

%\input{sect_summary.tex}
%%%%%%%%%%%%%%%%%%%%%%%%%%%%%%%%%%%%%%%%%%%%%%%%%%%%%%%%%%%%%%%%%%
\section{Summary}
\label{sect_summary}
%%%%%%%%%%%%%%%%%%%%%%%%%%%%%%%%%%%%%%%%%%%%%%%%%%%%%%%%%%%%%%%%%%

This paper discusses the potential of precision searches 
in dijet invariant masses at the HL-LHC and  HE-LHC.
It was illustrated 
that the HE-LHC provides a significantly higher reach for dijet searches than the HL-LHC, even
for rather modest $100~\fbb$ luminosity. 
We provide the relevant 95\% C.L. upper limits obtained from the
$\mjj$  distribution on cross-section times the  branching ratio for BSM models 
predicting heavy particles decaying to two jets, 
including jets identified as $b$-jets.
The limits at particle level were obtained for signals approximated with Gaussian distributions, as well as for 
$Z'$ signal shapes created using \pythia MC.

The dijet masses in semi-inclusive events with associated leptons provides particularly interesting data for searches, 
since they can be well measured in a large range of dijet masses without biases 
from triggers,  
and are less affected by the large rate of inclusive jet events.
We have illustrated that a data-driven determination of the shape background at the HL-LHC and  HE-LHC
should be performed with the relative statistical precision
of  $0.01\%$ per data point for $\mjj<1$~TeV.
It was shown that a requirement  to observe an isolated muon 
increases sensitivity to vector-boson and top-quark production.
With the expected statistical precision for the $\mjj$ measurements,  
the HL-LHC and  HE-LHC experiments  
will be sensitive to the shape of the $\mjj$ distribution of these processes.

\pythia MC simulations were used for testing the  data-driven approach used at the LHC for the  extraction of BSM signals.
When the fit function is applied to the  $\mjj$ distribution that matches  
the projected HL-LHC luminosity, 
we observe biases that limit detection of peaks in the $\mjj$ distributions.
Such biases are due to more complex shapes of 
the $\mjj$ distributions in \pythia, 
reflecting the underlying partonic kinematics, compared to the analytic function used at the LHC in the past.

\section*{Acknowledgments}

We thanks G.~Bodwin, S.~Mrenna and T.~Sj\"{o}strand for the discussion of the fit function used in this analysis.
The submitted manuscript has been created by UChicago Argonne, LLC,
Operator of Argonne National Laboratory (``Argonne'').
Argonne, a U.S. Department of Energy Office of Science laboratory,
is operated under Contract No. DE-AC02-06CH11357.
This research was made possible by an allocation of computing time through the ASCR Leadership Computing Challenge (ALCC) program. This research used resources of the National Energy Research Scientific Computing Center, a DOE Office of Science User Facility supported by the Office of Science of the U.S. Department of Energy under Contract No. DE-AC02-05CH11231.

%%%%%%%%%%%%%%%%%%%%%% references %%%%%%%%%%%%%%%%%%%%%%%%%%%%%%
\newpage
\clearpage
\section*{References}
\bibliography{biblio}

\end{document}